# HAT-P-44b, HAT-P-45b, AND HAT-P-46b: THREE TRANSITING HOT JUPITERS IN POSSIBLE MULTI-PLANET SYSTEMS[†]

J. D. Hartman[1], G. Á. Bakos[1,*,+], G. Torres[2], G. Kovács[3,15], J. A. Johnson[4], A. W. Howard[5], G. W. Marcy[6], D. W. Latham[2], A. Bieryla[2], L. A. Buchhave[7], W. Bhatti[1], B. Béky[2], Z. Csubry[1], K. Penev[1], M. de Val-Borro[1], R. W. Noyes[2], D. A. Fischer[8], G. A. Esquerdo[2], M. Everett[9], T. Szklenár[10], G. Zhou[11], D. Bayliss[11], A. Shporer[12,13,4], B. J. Fulton[5], R. Sanchis-Ojeda[14], E. Falco[2], J. Lázár[10], I. Papp[10], P. Sári[10]

## ABSTRACT

We report the discovery by the HATNet survey of three new transiting extrasolar planets orbiting moderately bright (V=13.2, 12.8 and 11.9) stars. The planets have orbital periods of 4.3012, 3.1290, and 4.4631 days, masses of 0.39, 0.89, and 0.49 $M_J$, and radii of 1.28, 1.43, and 1.28 $R_J$. The stellar hosts have masses of 0.94, 1.26, and 1.28 $M_\odot$. Each system shows significant systematic variations in its residual radial velocities indicating the possible presence of additional components. Based on its Bayesian evidence, the preferred model for HAT-P-44 consists of two planets, including the transiting component, with the outer planet having a period of 220 d and a minimum mass of 1.6 $M_J$. Due to aliasing we cannot rule out an alternative solution for the outer planet having a period of 438 d and a minimum mass of 3.7 $M_J$. For HAT-P-45 at present there is not enough data to justify the additional free parameters included in a multi-planet model, in this case a single-planet solution is preferred, but the required jitter of $22.5 \pm 6.3\,\mathrm{m\,s^{-1}}$ is relatively high for a star of this type. For HAT-P-46 the preferred solution includes a second planet having a period of 78 d and a minimum mass of 2.0 $M_J$, however the preference for this model over a single-planet model is not very strong. While substantial uncertainties remain as to the presence and/or properties of the outer planetary companions in these systems, the inner transiting planets are well characterized with measured properties that are fairly robust against changes in the assumed models for the outer planets. Continued RV monitoring is necessary to fully characterize these three planetary systems, the properties of which may have important implications for understanding the formation of hot Jupiters.

*Keywords:* planetary systems — stars: individual (HAT-P-44, GSC 3465-00123, HAT-P-45, GSC 5102-00262, HAT-P-46, GSC 5100-00045) — techniques: spectroscopic, photometric

## 1. INTRODUCTION

There is mounting evidence that systems containing close-in, gas-giant planets (hot Jupiters) are fundamentally different from systems that do not contain such a planet. These differences are seen in the occurrence rate of multiple planets between systems with and without hot Jupiters and in the distribution of projected orbital obliquities[19] of hot Jupiters compared to that of other planets.

Out of the 187 systems listed in the exoplanets orbit database[20] (Wright et al. 2011) containing a planet with $P < 10$ d and $M_p \sin i > 0.1\,M_J$, only 5 (2.7%) include confirmed, and well-characterized outer planets (these are $\upsilon$ And, Butler et al. 1997, 1999; HD 217107, Fischer et al. 1999, Vogt et al. 2005; HD 187123, Butler et al. 1998, Wright et al. 2007; HIP 14810, Wright et al. 2007; and HAT-P-13, Bakos et al. 2009). By contrast there are 87 multi-planet systems among the 395 systems (22%) in the database that do not have a hot Jupiter. In addition to the 5 confirmed multi-planet hot Jupiter systems, there are a number of other hot-Jupiter-bearing systems for which long term trends in their RVs have been reported. These trends could be due to long-period planetary companions, but their periods are significantly longer than the time spanned by the observations, and one cannot generally rule out stellar mass com-panions (a few examples from the Hungarian Automated Telescope Network, or HATNet, survey include HAT-P-7, Pál et al. 2008; HAT-P-17, Howard et al. 2012, Fulton et al. 2013; HAT-P-19, Hartman et al. 2011; and HAT-P-34, Bakos et al. 2012). Differences in the occurrence rate of multiple planets between hot-Jupiter-hosting systems and other systems are also apparent from the sample of *Kepler* transiting planet candidates (Latham et al. 2011).

[1] Department of Astrophysical Sciences, Princeton University, Princeton, NJ 08544; email: jhartman@astro.princeton.edu
[*] Sloan Fellow
[+] Packard Fellow
[2] Harvard-Smithsonian Center for Astrophysics, Cambridge, MA
[3] Konkoly Observatory, Budapest, Hungary
[4] Department of Astrophysics, California Institute of Technology, Pasadena, CA
[5] Institute for Astronomy, University of Hawaii, Honolulu, HI 96822
[6] Department of Astronomy, University of California, Berkeley, CA
[7] Niels Bohr Institute, Copenhagen University, Denmark
[8] Department of Astronomy, Yale University, New Haven, CT
[9] National Optical Astronomy Observatory, 950 N. Cherry Ave., Tucson, AZ 85719, USA
[10] Hungarian Astronomical Association, Budapest, Hungary
[11] Research School of Astronomy and Astrophysics, Australian National University, Canberra, ACT 2611, Australia
[12] Las Cumbres Observatory Global Telescope Network, 6740 Cortona Drive, Suite 102, Santa Barbara, CA 93117, USA
[13] Department of Physics, Broida Hall, University of California, Santa Barbara, CA 93106, USA
[14] Kavli Institute for Astrophysics and Space Research, Massachusetts Institute of Technology, Cambridge, MA
[15] Department of Physics and Astrophysics, University of North Dakota, Grand Forks, ND, USA

[†] Based in part on observations obtained at the W. M. Keck Observatory, which is operated by the University of California and the California Institute of Technology. Keck time has been granted by NOAO (A284Hr) and NASA (N154Hr, N108Hr).

[19] We use the term obliquity here to refer to the angle between the orbital axis of a planet and the spin axis of its host star.
[20] exoplanets.org, accessed 04 June 2013.

Observations of the Rossiter-McLaughlin effect have revealed that hot Jupiters exhibit a broad range of projected obliquities (e.g. Albrecht et al. 2012). In contrast, the $\sim 5$ multi-planet systems, not containing a hot Jupiter, for which the projected obliquity of at least one of the planets has been determined, are all aligned (Albrecht et al. 2013). Differences in the obliquities have been interpreted as indicating different migration mechanisms between the two populations (Sanchis-Ojeda et al. 2012; Albrecht et al. 2013).

There are, however, selection effects which complicate this picture. While most multi-planet systems have been discovered by RV surveys or by the NASA *Kepler* space mission, the great majority of hot Jupiters have been discovered by ground-based transiting planet searches. For these latter surveys access to high-precision RV resources may be scarce, and the candidates are usually several magnitudes fainter than those targeted by RV surveys. To deal with these factors, ground-based transit surveys leverage the known ephemerides of their candidates so as to minimize the number of RV observations needed to detect the orbital variation. In practice this means that many published hot Jupiters do not have the long-term RV monitoring that would be necessary to detect other planetary companions, if present. Moreover, ground-based surveys produce light curves with much shorter time coverage and poorer precision than *Kepler*, so whereas *Kepler* has identified numerous multi-transiting-planet systems, ground-based surveys have not yet discovered any such systems.

In this paper we report the discovery of three new transiting planet systems by the HATNet survey (Bakos et al. 2004). The transiting planets are all classical hot Jupiters, confirmed through a combination of ground-based photometry and spectroscopy, including high-precision radial velocity (RV) measurements made with Keck-I/HIRES which reveal the orbital motion of the star about the planet–star center-of-mass. In addition to the orbital motion due to the transiting planets, the RV measurements for all three systems show systematic variations indicating the possible presence of additional planetary-mass components. As we will show, for two of these systems (HAT-P-44 and HAT-P-46) we find that the observations are best explained by multi-planet models, while for the third system (HAT-P-45) additional RV observations would be necessary to claim an additional planet.

In Section 2 we summarize the detection of the photometric transit signal and the subsequent spectroscopic and photometric observations of each star to confirm the planets. In Section 3 we analyze the data to determine the stellar and planetary parameters. Our findings are discussed in Section 4.

## 2. OBSERVATIONS

The observational procedure employed by HATNet to discover Transiting Extrasolar Planets (TEPs) has been described in several previous discovery papers (e.g. Bakos et al. 2010; Latham et al. 2009). In the following subsections we highlight specific details of this procedure that are relevant to the discoveries presented in this paper.

### 2.1. *Photometric detection*

Table 1 summarizes the HATNet discovery observations of each new planetary system. The HATNet images were processed and reduced to trend-filtered light curves following the procedure described by Bakos et al. (2010). The light curves were searched for periodic box-shaped signals using the Box Least-Squares (BLS; see Kovács et al. 2002) method. Figure 1 shows phase-folded HATNet light curves for HAT-P-44, HAT-P-45, and HAT-P-46 which were selected as showing highly significant transit signals based on their BLS spectra. Cross-identifications, positions, and the available photometry on an absolute scale are provided later in the paper together with other system parameters (Table 10).

We removed the detected transits from the HATNet light curves for each of these systems and searched the residuals for additional transits using BLS, and for other periodic signals using the Discrete Fourier Transform (DFT). Using DFT we do not find a significant signal in the frequency range $0\,\mathrm{d}^{-1}$ to $50\,\mathrm{d}^{-1}$ in the light curves of any of these systems. For HAT-P-44 we exclude signals with amplitudes above 1.2 mmag, for HAT-P-45 we exclude signals with amplitudes above 1.1 mmag, and for HAT-P-46 we exclude signals with amplitudes above 0.6 mmag. Similarly we do not detect additional transit signals in the light curves of HAT-P-44 or HAT-P-45. For HAT-P-46 we do detect a marginally significant transit signal with a short period of $P = 0.388\,\mathrm{d}$, a depth of 2.3 mmag, and a S/N in the BLS spectrum of 8.5. The period is neither a harmonic nor an alias of the primary transit signal. Based on our prior experience following up similar signals detected in HATNet light curves we consider this likely to be a false alarm, but mention it here for full disclosure.

### 2.2. *Reconnaissance Spectroscopy*

High-resolution, low-S/N "reconnaissance" spectra were obtained for HAT-P-44, HAT-P-45, and HAT-P-46 using the Tillinghast Reflector Echelle Spectrograph (TRES; Fűresz 2008) on the 1.5 m Tillinghast Reflector at FLWO. Medium-resolution reconnaissance spectra were also obtained for HAT-P-45 and HAT-P-46 using the Wide Field Spectrograph (WiFeS) on the ANU 2.3 m telescope at Siding Spring Observatory. The reconnaissance spectroscopic observations and results for each system are summarized in Table 2. The TRES observations were reduced and analyzed following the procedure described by Quinn et al. (2012); Buchhave et al. (2010), yielding RVs with a precision of $\sim 50\,\mathrm{m\,s}^{-1}$, and an absolute velocity zeropoint accuracy of $\sim 100\,\mathrm{m\,s}^{-1}$. The WiFeS observations were reduced and analyzed as described in Bayliss et al. (2013), providing RVs with a precision of $2.8\,\mathrm{km\,s}^{-1}$.

Based on the observations summarized in Table 2 we find that all three systems have RMS residuals consistent with no significant RV variation within the precision of the measurements (the WiFeS observations of HAT-P-46 have an RMS of $3.3\,\mathrm{km\,s^{-1}}$ which is only slightly above the precision determined from observations of RV stable stars). All spectra were single-lined, i.e., there is no evidence that any of these targets consist of more than one star. The gravities for all of the stars indicate that they are dwarfs.

**Table 1**
Summary of photometric observations

| Instrument/Field | Date(s) | Number of Images | Cadence (sec) | Filter |
|---|---|---|---|---|
| **HAT-P-44** | | | | |
| HAT-5/G145 | 2006 Jan–2006 Jul | 2880 | 330 | $I$ band |
| HAT-6/G146 | 2010 Apr–2010 Jul | 6668 | 210 | $r$ band |
| KeplerCam | 2011 Mar 19 | 112 | 134 | $i$ band |
| BOS | 2011 Apr 14 | 176 | 131 | $r$ band |
| KeplerCam | 2011 Apr 14 | 85 | 134 | $i$ band |
| KeplerCam | 2011 May 27 | 176 | 134 | $i$ band |
| **HAT-P-45** | | | | |
| HAT-5/G432 | 2010 Sep–2010 Oct | 272 | 330 | $r$ band |
| HAT-8/G432 | 2010 Apr–2010 Oct | 7309 | 210 | $r$ band |
| KeplerCam | 2011 Apr 02 | 133 | 73 | $i$ band |
| KeplerCam | 2011 Apr 05 | 44 | 103 | $i$ band |
| FTN | 2011 Apr 30 | 197 | 50 | $i$ band |
| KeplerCam | 2011 May 22 | 174 | 64 | $i$ band |
| KeplerCam | 2011 Jun 10 | 146 | 64 | $i$ band |
| KeplerCam | 2011 Jul 05 | 99 | 103 | $i$ band |
| KeplerCam[a] | 2013 May 20 | 229 | 50 | $g$ band |
| **HAT-P-46** | | | | |
| HAT-5/G432 | 2010 Sep–2010 Oct | 300 | 330 | $r$ band |
| HAT-8/G432 | 2010 Apr–2010 Oct | 7633 | 210 | $r$ band |
| KeplerCam | 2011 May 05 | 392 | 44 | $i$ band |
| KeplerCam | 2011 May 14 | 368 | 49 | $i$ band |
| KeplerCam | 2011 May 23 | 247 | 39 | $i$ band |

[a] This observation was included in the blend analysis of the system, but was not included in the analysis conducted to determine the system parameters.

### 2.3. *High resolution, high S/N spectroscopy*

We obtained high-resolution, high-S/N spectra of each of these objects using HIRES (Vogt et al. 1994) on the Keck-I telescope in Hawaii. The data were reduced to radial velocities in the barycentric frame following the procedure described by Butler et al. (1996). The RV measurements and uncertainties are given in Tables 3-5 for HAT-P-44 through HAT-P-46, respectively. The period-folded data, along with our best fit described below in Section 3 are displayed in Figures 2-4.

We also show the chromospheric activity $S$ index and the spectral line bisector spans. The $S$ index for each star was computed following Isaacson & Fischer (2010) and converted to $\log R'_{\mathrm{HK}}$ following Noyes et al. (1984). We find median values of $\log R'_{\mathrm{HK}} = -5.247$, $-5.394$, and $-5.257$ for HAT-P-44 through HAT-P-46, respectively. These values imply that all three stars are chromospherically quiet. The bisector spans were computed as in Torres et al. (2007) and Bakos et al. (2007) and show no detectable variation in phase with the RVs, allowing us to rule out various blend scenarios as possible explanations of the observations (see Section 3.2).

### 2.4. *Photometric follow-up observations*

In order to permit a more accurate modeling of the light curves, we conducted additional photometric observations of each of the transiting planet systems. For this purpose we made use of the KeplerCam CCD camera on the FLWO 1.2 m telescope, the CCD imager on the 0.8 m remotely operated Byrne Observatory at Sedgwick (BOS) reserve in California, and the Spectral Instrument CCD on the 2.0 m Faulkes Telescope North (FTN) at Haleakala Observatory in Hawaii. Both BOS and FTN are operated by the Las Cumbres Observatory Global Telescope (LCOGT; Brown et al. 2013). The observations for each target are summarized in Table 1.

The reduction of the KeplerCam images was performed as described by Bakos et al. (2010). The BOS and FTN observations were reduced in a similar manner. The resulting differential light curves were further filtered using the External Parameter Decorrelation (EPD) and Trend Filtering Algorithm (TFA)[21] methods applied simultaneously with the light curve modeling so that uncertainties in the noise filtering process contribute to the uncertainties on the physical parameters (for more details, see Bakos et al. 2010). The final time series, together with our best-fit transit light curve model, are shown in the top portion of Figures 5-7 for HAT-P-44 through HAT-P-46, respectively; the individual measurements are reported in Tables 6-8.

## 3. ANALYSIS

### 3.1. *Properties of the parent star*

[21] EPD and TFA both involve fitting the light curve as a linear combination of trend basis vectors. The EPD vectors are a set of light curve specific signals, such as the hour angle of the observations and the Full Width at Half Maximum (FWHM) of the Point Spread Function (PSF). The TFA vectors are the differential light curves of a carefully selected sample of comparison stars in the same field of view as the target.

**Table 2**
Summary of reconnaissance spectroscopy observations

| Instrument | $HJD - 2400000$ | $T_{\rm eff\star}$[a] (K) | $\log g_\star$ (cgs) | $v \sin i$ ($\rm km\,s^{-1}$) | RV ($\rm km\,s^{-1}$) |
|---|---|---|---|---|---|
| **HAT-P-44** | | | | | |
| TRES | 55557.01323 | 5250 | 4.5 | 2 | -34.042 |
| TRES | 55583.91926 | 5250 | 4.5 | 2 | -34.047 |
| **HAT-P-45** | | | | | |
| WiFeS | 55646.25535 | ··· | ··· | ··· | 18.9 |
| WiFeS | 55648.19634 | ··· | ··· | ··· | 16.6 |
| WiFeS | 55649.24624 | ··· | ··· | ··· | 18.5 |
| WiFeS | 55666.31876 | ··· | ··· | ··· | 20.1 |
| TRES | 55691.96193 | 6500 | 4.5 | 10 | 23.162 |
| **HAT-P-46** | | | | | |
| WiFeS | 55644.28771 | ··· | ··· | ··· | -21.1 |
| WiFeS | 55646.25316 | ··· | ··· | ··· | -29.6 |
| WiFeS | 55647.21574 | ··· | ··· | ··· | -21.3 |
| WiFeS | 55647.21882 | ··· | ··· | ··· | -21.7 |
| WiFeS | 55648.17221 | ··· | ··· | ··· | -23.9 |
| WiFeS | 55649.21348 | ··· | ··· | ··· | -25.0 |
| TRES | 55659.92299 | 6000 | 4.0 | 6 | -21.314 |
| TRES | 55728.82463 | 6000 | 4.0 | 6 | -21.385 |

[a] The stellar parameters listed for the TRES observations are the parameters of the theoretical template spectrum used to determine the velocity from the Mg b order. These parameters assume solar metallicity.

**Table 3**
Relative radial velocities, bisector spans, and activity index measurements of HAT-P-44.

| BJD (2,454,000+) | RV[a] ($\rm m\,s^{-1}$) | $\sigma_{\rm RV}$[b] ($\rm m\,s^{-1}$) | BS ($\rm m\,s^{-1}$) | $\sigma_{\rm BS}$ ($\rm m\,s^{-1}$) | S[c] | Phase |
|---|---|---|---|---|---|---|
| 1606.11852 | 103.86 | 4.35 | 4.86 | 6.27 | 0.1310 | 0.885 |
| 1612.12701 | −5.37 | 3.65 | −6.08 | 4.51 | 0.1220 | 0.282 |
| 1612.14300 | ··· | ··· | 19.68 | 8.23 | 0.1340 | 0.286 |
| 1613.03742 | 67.53 | 12.22 | ··· | ··· | 0.2320 | 0.494 |
| 1614.11905 | 99.23 | 4.03 | −0.94 | 5.65 | 0.1060 | 0.745 |
| 1634.02686 | 4.99 | 3.88 | 0.70 | 5.06 | 0.1460 | 0.374 |
| 1663.99061 | −13.54 | 3.76 | −1.47 | 8.11 | 0.1270 | 0.340 |
| 1671.88450 | −25.58 | 4.06 | −6.52 | 5.82 | 0.1150 | 0.176 |
| 1672.95630 | 2.88 | 4.01 | −8.64 | 6.49 | 0.1340 | 0.425 |
| 1673.87329 | 57.51 | 3.84 | 13.66 | 6.67 | 0.1230 | 0.638 |
| 1696.81647 | 7.20 | 4.38 | 28.16 | 19.64 | 0.0878 | 0.972 |
| 1697.81428 | −47.86 | 3.61 | 16.33 | 16.80 | 0.1130 | 0.204 |
| 1698.88909 | ··· | ··· | ··· | ··· | 0.2110 | 0.454 |
| 1699.79550 | 17.57 | 3.65 | 23.87 | 12.27 | 0.1410 | 0.665 |
| 1700.84720 | −1.91 | 3.34 | 4.25 | 4.48 | 0.1180 | 0.909 |
| 1703.83608 | −14.10 | 3.34 | 6.33 | 6.88 | 0.1230 | 0.604 |
| 1704.77501 | 11.45 | 3.83 | −12.18 | 10.70 | 0.1230 | 0.822 |
| 1705.79848 | −57.55 | 3.40 | −1.17 | 5.64 | 0.0890 | 0.060 |
| 1707.82993 | −27.03 | 3.90 | 0.26 | 5.30 | 0.1370 | 0.533 |
| 1723.83103 | −105.67 | 4.37 | −9.10 | 4.26 | 0.1330 | 0.253 |
| 1945.04563 | −11.27 | 4.68 | ··· | ··· | ··· | 0.683 |
| 2019.98862 | −38.87 | 4.18 | ··· | ··· | ··· | 0.107 |
| 2115.75933 | −27.58 | 3.90 | ··· | ··· | ··· | 0.373 |

**Note**. — Note that for the iodine-free template exposures we do not measure the RV but do measure the BS and S index. Such template exposures can be distinguished by the missing RV value.

[a] The zero-point of these velocities is arbitrary. An overall offset $\gamma_{\rm rel}$ fitted to these velocities in Section 3.4 has *not* been subtracted.

[b] Internal errors excluding the component of astrophysical jitter considered in Section 3.4. The formal errors are likely underestimated in cases where $\sigma_{\rm RV} > 10\,\rm m\,s^{-1}$, as the HIRES Doppler code is not reliable for low S/N observations.

[c] Chromospheric activity index computed as in Isaacson & Fischer (2010).

**Table 4**
Relative radial velocities, bisector spans, and activity index measurements of HAT-P-45.

| BJD (2,454,000+) | RV[a] ($m\,s^{-1}$) | $\sigma_{RV}$[b] ($m\,s^{-1}$) | BS ($m\,s^{-1}$) | $\sigma_{BS}$ ($m\,s^{-1}$) | S[c] | Phase |
|---|---|---|---|---|---|---|
| 1668.07265 | ··· | ··· | −16.03 | 8.62 | 0.1220 | 0.213 |
| 1696.93207 | 5.66 | 7.42 | 21.27 | 9.92 | 0.1310 | 0.436 |
| 1697.91247 | ··· | ··· | 25.71 | 6.26 | 0.1260 | 0.750 |
| 1697.92558 | 123.23 | 8.08 | 18.73 | 11.98 | 0.1300 | 0.754 |
| 1698.91871 | −18.91 | 7.81 | −1.89 | 22.68 | 0.1120 | 0.071 |
| 1699.90750 | −74.19 | 6.89 | −2.21 | 17.77 | 0.1170 | 0.387 |
| 1701.10094 | 97.43 | 5.92 | −48.91 | 8.94 | 0.1040 | 0.769 |
| 1703.91095 | 57.59 | 6.41 | −18.46 | 11.91 | 0.1140 | 0.667 |
| 1705.88887 | −128.98 | 6.28 | −8.37 | 9.85 | 0.1200 | 0.299 |
| 1706.88594 | 69.84 | 6.70 | 4.78 | 9.59 | 0.1150 | 0.617 |
| 1853.73646 | 28.34 | 6.19 | ··· | ··· | ··· | 0.550 |
| 2020.03172 | 124.79 | 7.60 | ··· | ··· | ··· | 0.696 |

**Note**. — Note that for the iodine-free template exposures we do not measure the RV but do measure the BS and S index. Such template exposures can be distinguished by the missing RV value.

[a] The zero-point of these velocities is arbitrary. An overall offset $\gamma_{rel}$ fitted to these velocities in Section 3.4 has *not* been subtracted.

[b] Internal errors excluding the component of astrophysical jitter considered in Section 3.4.

[c] Chromospheric activity index computed as in Isaacson & Fischer (2010).

**Table 5**
Relative radial velocities, bisector spans, and activity index measurements of HAT-P-46.

| BJD (2,454,000+) | RV[a] ($m\,s^{-1}$) | $\sigma_{RV}$[b] ($m\,s^{-1}$) | BS ($m\,s^{-1}$) | $\sigma_{BS}$ ($m\,s^{-1}$) | S[c] | Phase |
|---|---|---|---|---|---|---|
| 1696.90831 | −3.49 | 5.78 | −3.77 | 5.30 | 0.1310 | 0.008 |
| 1696.91767 | ··· | ··· | 13.10 | 8.21 | 0.1280 | 0.010 |
| 1697.89794 | −60.41 | 5.22 | 3.55 | 4.74 | 0.1290 | 0.230 |
| 1698.90686 | −38.37 | 5.28 | −0.87 | 10.21 | 0.1510 | 0.456 |
| 1699.89562 | −0.43 | 4.90 | −1.79 | 3.95 | 0.1250 | 0.677 |
| 1703.89107 | −24.17 | 4.57 | 4.96 | 5.79 | 0.1260 | 0.572 |
| 1704.88237 | −3.19 | 5.33 | 2.79 | 8.88 | 0.1240 | 0.794 |
| 1705.87529 | −55.16 | 5.20 | −9.04 | 7.16 | 0.1270 | 0.017 |
| 1706.87230 | −115.33 | 5.57 | −1.42 | 3.72 | 0.1270 | 0.240 |
| 1853.71982 | −63.30 | 5.34 | ··· | ··· | ··· | 0.143 |
| 1997.09764 | 20.23 | 13.90 | ··· | ··· | ··· | 0.268 |
| 2113.77814 | −87.17 | 6.61 | ··· | ··· | ··· | 0.411 |
| 2195.72876 | 26.55 | 5.74 | ··· | ··· | ··· | 0.773 |

**Note**. — Note that for the iodine-free template exposures we do not measure the RV but do measure the BS and S index. Such template exposures can be distinguished by the missing RV value.

[a] The zero-point of these velocities is arbitrary. An overall offset $\gamma_{rel}$ fitted to these velocities in Section 3.4 has *not* been subtracted.

[b] Internal errors excluding the component of astrophysical jitter considered in Section 3.4. The formal errors are likely underestimated in cases where $\sigma_{RV} > 10\,m\,s^{-1}$, as the HIRES Doppler code is not reliable for low S/N observations.

[c] Chromospheric activity index computed as in Isaacson & Fischer (2010).

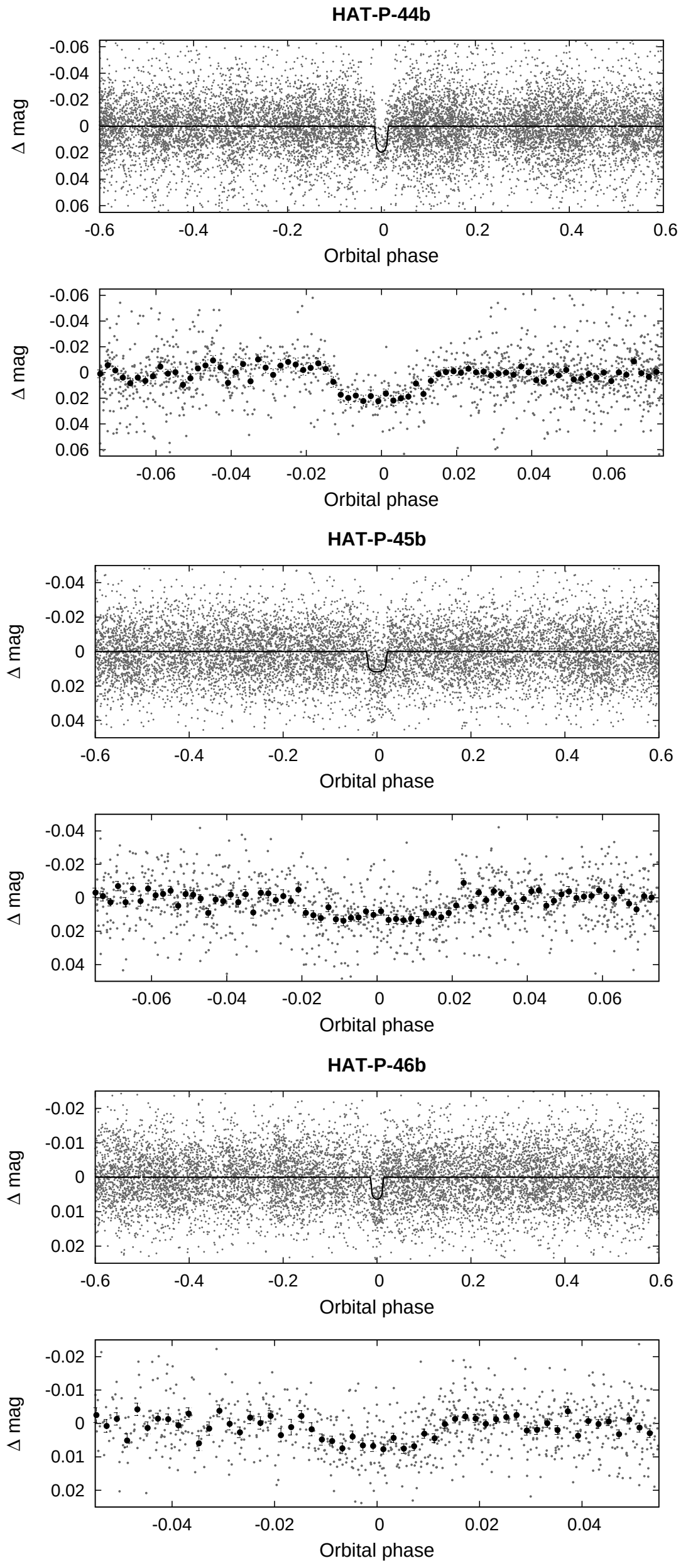

**Figure 1.** HATNet light curves of HAT-P-44 (top), HAT-P-45 (middle) and HAT-P-46 (bottom). See Table 1 for a summary of the observations. For each planet we show two panels. The top panel shows the unbinned light curve folded with the period resulting from the global fit described in Section 3. The solid line shows the model fit to the light curve (Section 3.4). The bottom panel shows the region zoomed-in on the transit. The dark filled circles show the light curve binned in phase with a bin size of 0.002. These are plotted with $1\sigma$ uncertainties.

**Table 6**
High-precision differential photometry of HAT-P-44.

| BJD (2,400,000+) | Mag[a] | $\sigma_{\rm Mag}$ | Mag(orig)[b] | Filter |
|---|---|---|---|---|
| 55640.84396 | −0.00143 | 0.00122 | 12.01460 | $i$ |
| 55640.84597 | 0.00034 | 0.00119 | 12.01570 | $i$ |
| 55640.84767 | −0.00081 | 0.00119 | 12.01670 | $i$ |
| 55640.84922 | 0.00007 | 0.00122 | 12.01650 | $i$ |
| 55640.85109 | −0.00137 | 0.00119 | 12.01450 | $i$ |
| 55640.85271 | 0.00200 | 0.00122 | 12.01870 | $i$ |
| 55640.85427 | −0.00050 | 0.00121 | 12.01580 | $i$ |
| 55640.85593 | −0.00410 | 0.00122 | 12.01310 | $i$ |
| 55640.85747 | −0.00288 | 0.00122 | 12.01370 | $i$ |
| 55640.85902 | −0.00104 | 0.00119 | 12.01550 | $i$ |

**Note**. — This table is available in a machine-readable form in the online journal. A portion is shown here for guidance regarding its form and content.
[a] The out-of-transit level has been subtracted. These magnitudes have been subjected to the EPD and TFA procedures, carried out simultaneously with the transit fit.
[b] Raw magnitude values without application of the EPD and TFA procedures.

**Table 7**
High-precision differential photometry of HAT-P-45.

| BJD (2,400,000+) | Mag[a] | $\sigma_{\rm Mag}$ | Mag(orig)[b] | Filter |
|---|---|---|---|---|
| 55654.89724 | 0.01314 | 0.00112 | 11.34410 | $i$ |
| 55654.89809 | 0.01315 | 0.00112 | 11.34380 | $i$ |
| 55654.89894 | 0.01560 | 0.00112 | 11.34390 | $i$ |
| 55654.89981 | 0.01522 | 0.00112 | 11.34580 | $i$ |
| 55654.90065 | 0.01456 | 0.00113 | 11.34380 | $i$ |
| 55654.90152 | 0.01758 | 0.00112 | 11.34740 | $i$ |
| 55654.90235 | 0.01425 | 0.00112 | 11.34350 | $i$ |
| 55654.90322 | 0.01632 | 0.00112 | 11.34700 | $i$ |
| 55654.90407 | 0.01552 | 0.00113 | 11.34650 | $i$ |
| 55654.90491 | 0.01267 | 0.00112 | 11.34330 | $i$ |

**Note**. — This table is available in a machine-readable form in the online journal. A portion is shown here for guidance regarding its form and content.
[a] The out-of-transit level has been subtracted. These magnitudes have been subjected to the EPD and TFA procedures, carried out simultaneously with the transit fit.
[b] Raw magnitude values without application of the EPD and TFA procedures.

Stellar atmospheric parameters for each star were measured using our template spectra obtained with the Keck/HIRES instrument, and the analysis package known as Spectroscopy Made Easy (SME; Valenti & Piskunov 1996), along with the atomic line database of Valenti & Fischer (2005). For each star, SME yielded the following *initial* values and uncertainties:

- *HAT-P-44* – effective temperature $T_{\rm eff\star} = 5295 \pm 100$ K, metallicity [Fe/H] $= 0.33 \pm 0.1$ dex, stellar surface gravity $\log g_\star = 4.42 \pm 0.1$ (cgs), and projected rotational velocity $v \sin i = 0.2 \pm 0.5\,{\rm km\,s^{-1}}$.
- *HAT-P-45* – effective temperature $T_{\rm eff\star} = 6270 \pm 100$ K, metallicity [Fe/H] $= 0.03 \pm 0.1$ dex, stellar surface gravity $\log g_\star = 4.26 \pm 0.1$ (cgs), and projected rotational velocity $v \sin i = 9.0 \pm 0.5\,{\rm km\,s^{-1}}$.
- *HAT-P-46* – effective temperature $T_{\rm eff\star} = 6280 \pm 100$ K, metallicity [Fe/H] $= 0.38 \pm 0.1$ dex, stellar surface gravity $\log g_\star = 4.38 \pm 0.1$ (cgs), and projected rotational velocity $v \sin i = 4.5 \pm 0.5\,{\rm km\,s^{-1}}$.

These values were used to determine initial values for the limb-darkening coefficients, which we fix during the light curve modeling (Section 3.4). This modeling, when combined with theoretical stellar evolution models taken from the Yonsei-Yale (YY) series by Yi et al. (2001), provides a refined determination of the stellar surface gravity (Sozzetti et al. 2007) which we then fix in a second SME analysis of the spectra yielding our adopted atmospheric parameters. For HAT-P-44 the revised surface gravity is close enough to the initial SME value that we do not conduct a second SME analysis. The final adopted values of $T_{\rm eff\star}$, [Fe/H] and $v \sin i$ are listed for each star in Table 10. The values of $\log g_\star$, as well as of properties inferred from the evolution models (such as the stellar masses and radii) depend on the eccentricity and semi-amplitude of the transiting planet's orbit, which in turn depend on how the RV data are modeled. In modeling these data we varied the number of planets considered for a given system, and whether or not these planets are fixed to circular orbits. Although $T_{\rm eff\star}$, [Fe/H], and $v \sin i$ will also depend on the fixed value of $\log g_\star$ we found generally that $\log g_\star$ did not change enough between the models that provide a good fit to the data to justify carrying out a separate SME analysis using the $\log g_\star$ value determined from each model. As we discuss in Section 3.4.2 we tested numerous models; our final adopted values for these model-dependent parameters are presented in that section.

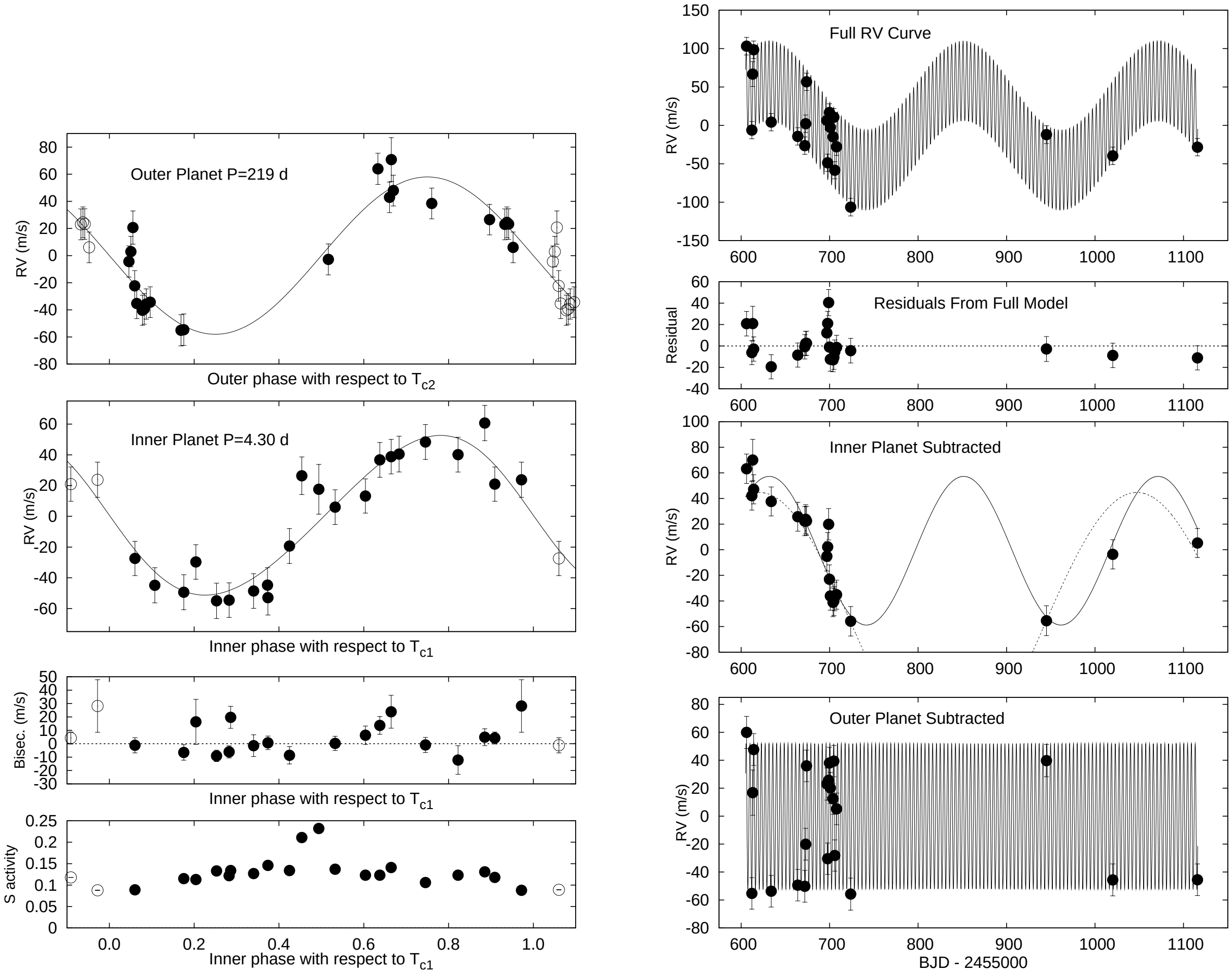

**Figure 2.** *Top left:* Keck/HIRES RV measurements for HAT-P-44 shown as a function of orbital phase for the outer planet HAT-P-44c assuming a period of $P = 220$ d, after subtracting off the model orbit for the inner transiting planet HAT-P-44b. Overplotted is our best-fit model (see Table 11). Zero phase corresponds to the time of mid-transit (assuming an edge on orbit; transits of this planet have not been detected). The center-of-mass velocity has been subtracted. Observations shown twice are plotted with open circles. *Second left:* Same as the top left panel, here we show the observations phased using the ephemeris of the transiting planet HAT-P-44b, after subtracting off Keplerian orbit due to HAT-P-44c. *Third left:* Bisector spans (BS), with the mean value subtracted, phased using the ephemeris for the transiting planet. The measurement from the template spectrum is included (see Section 3.2). *Bottom left:* Chromospheric activity index $S$ measured from the Keck spectra, phased using the ephemeris for the transiting planet. *Top right:* RV curve of HAT-P-44 as a function of time, together with our best-fit two planet model. *Second right:* RV residuals from the best-fit two-planet model as a function of time. *Third right:* RV curve of HAT-P-44 as a function of time after subtracting the orbit due to the inner planet HAT-P-44b. For reference, the dashed line in this plot shows the model for the $P = 438$ d alias signal. *Bottom right:* RV curve of HAT-P-44 as a function of time after subtracting the orbit due to the outer planet HAT-P-44c. Note the difference in vertical scales for all panels. RV uncertainties in this figure include the jitter which is estimated during the fitting procedure (see Section 3.4.1) and added in quadrature to the formal uncertainties.

The inferred location of each star in a diagram of $a/R_\star$ versus $T_{\rm eff\star}$, analogous to the classical H-R diagram, is shown in Figure 8. In each case the stellar properties and their $1\sigma$ and $2\sigma$ confidence ellipsoids are displayed against the backdrop of model isochrones for a range of ages, and the appropriate stellar metallicity. For comparison, the locations implied by the initial SME results for HAT-P-45 and HAT-P-46 are also shown (in each case with a triangle).

We determine the distance and extinction to each star by comparing the $J$, $H$ and $K_S$ magnitudes from the 2MASS Catalogue (Skrutskie et al. 2006), and the $V$ and $I_C$ magnitudes from the TASS Mark IV Catalogue (Droege et al. 2006), to the expected magnitudes

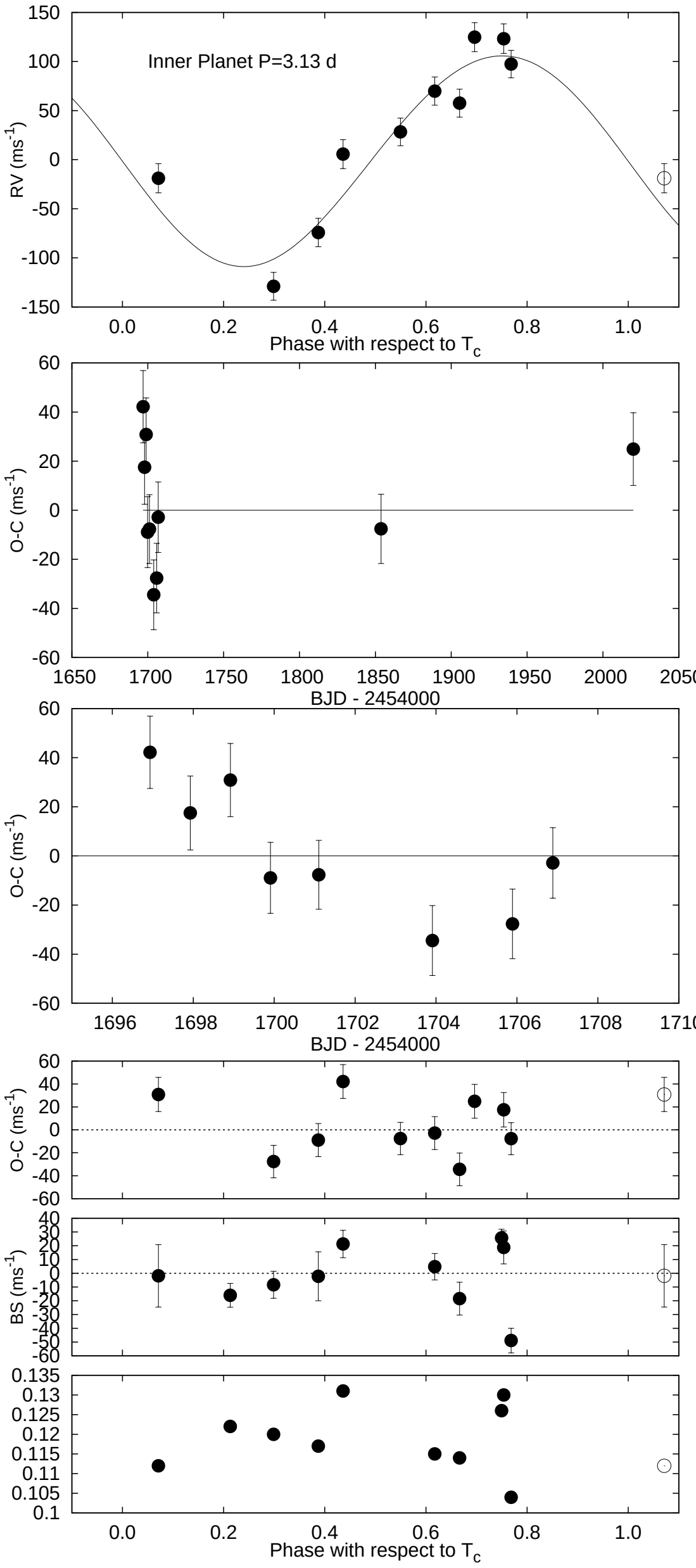

**Figure 3.** *Top panel:* Keck/HIRES RV measurements for HAT-P-45 shown as a function of orbital phase, along with our best-fit single-planet model (see Table 13). Zero phase corresponds to the time of mid-transit. The center-of-mass velocity has been subtracted. *Second panel:* Velocity $O-C$ residuals from the best fit shown as a function of time. *Third panel:* Same as the second panel, here we zoom-in on the residuals for the first 8 observations. *Fourth panel:* Velocity *O–C* residuals shown as a function of orbital phase. *Fifth panel:* Bisector spans (BS), with the mean value subtracted. The measurement from the template spectrum is included (see Section 3.2). *Bottom panel:* Chromospheric activity index $S$ measured from the Keck spectra. Note the different vertical scales of the panels. Observations shown twice are represented with open symbols.

from the stellar models. We use the transformations by Carpenter (2001) to convert the 2MASS magnitudes to the photometric system of the models (ESO), and use the Cardelli et al. (1989) extinction law, assuming a total-to-selective extinction ratio of $R_V = 3.1$, to relate the extinction in each band-pass to the $V$-band extinction $A_V$. The resulting $A_V$ and distance measurements are

**Table 8**
High-precision differential photometry of HAT-P-46.

| BJD (2,400,000+) | Mag[a] | $\sigma_{\rm Mag}$ | Mag(orig)[b] | Filter |
|---|---|---|---|---|
| 55687.79447 | −0.00440 | 0.00112 | 10.45700 | $i$ |
| 55687.79497 | 0.00288 | 0.00113 | 10.46460 | $i$ |
| 55687.79596 | −0.00087 | 0.00113 | 10.46190 | $i$ |
| 55687.79647 | −0.00065 | 0.00113 | 10.46210 | $i$ |
| 55687.79699 | −0.00327 | 0.00112 | 10.45850 | $i$ |
| 55687.79749 | 0.00222 | 0.00112 | 10.46330 | $i$ |
| 55687.79798 | 0.00089 | 0.00113 | 10.46300 | $i$ |
| 55687.79850 | −0.00339 | 0.00112 | 10.45840 | $i$ |
| 55687.79901 | 0.00378 | 0.00112 | 10.46610 | $i$ |
| 55687.79952 | 0.00308 | 0.00113 | 10.46470 | $i$ |

**Note.** — This table is available in a machine-readable form in the online journal. A portion is shown here for guidance regarding its form and content.
[a] The out-of-transit level has been subtracted. These magnitudes have been subjected to the EPD and TFA procedures, carried out simultaneously with the transit fit.
[b] Raw magnitude values without application of the EPD and TFA procedures.

given with the other model-dependent parameters. We find that HAT-P-44 is not significantly affected by extinction, consistent with the Schlegel et al. (1998) dust maps which yield a total extinction of $A_V = 0.038$ mag along the line of sight to HAT-P-44. HAT-P-45 and HAT-P-46, on the other hand, have low Galactic latitudes ($b = 6\fdg0$ and $b = 9\fdg6$, respectively), and are significantly affected by extinction. We find $A_V = 1.900 \pm 0.169$ mag and $A_V = 0.832 \pm 0.145$ mag for our preferred models for HAT-P-45 and HAT-P-46, respectively. For comparison, the Schlegel et al. (1998) maps yield a *total* line of sight extinction of $A_V = 5.89$ mag and $A_V = 3.25$ mag for HAT-P-45 and HAT-P-46, respectively, or $A_V \sim 0.8$ mag to both sources after applying the distance and excess extinction corrections given by Bonifacio et al. (2000). At these low Galactic latitudes the extinction estimates based on the Schlegel et al. (1998) dust maps are not reliable, so the discrepancy between the dust-map-based and photometry-based $A_V$ estimates for HAT-P-45 is not unexpected. After correcting for extinction the measured and expected photometric color indices are consistent for each star.

### 3.2. *Excluding Blend Scenarios*

To rule out the possibility that any of these objects might be a blended stellar eclipsing binary system we carried out a blend analysis as described in Hartman et al. (2012).

We find that for HAT-P-44 we can exclude most blend models, consisting either of a hierarchical triple star system, or a blend between a background eclipsing binary and a foreground bright star, based on the light curves. Those models that cannot be excluded with at least $5\sigma$ confidence would have been detected as obviously double-lined systems, showing many km s$^{-1}$ RV and BS variations.

For HAT-P-45 and HAT-P-46 the significant reddening (Section 3.1) allows a broader range of blend scenarios to fit the photometric data. For a system like HAT-P-44, where there is no significant reddening and the available calibrated broad-band photometry agrees well with the spectroscopically determined temperature, the cali-brated photometry places a strong constraint on blend scenarios where the two brightest stars in the blend have different temperatures. For HAT-P-45 and HAT-P-46, on the other hand, such blends can be accommodated by reducing the reddening in the fit. Indeed we find for both HAT-P-45 and HAT-P-46 that the calibrated broad-band photometry are fit slightly better by models that incorporate multiple stars (blends) together with reddening, than by a model consisting of only a single reddened star. The difference between these models is small enough, however, that we do not consider this improvement to be significant; such differences may be due to the true extinction law along this line of sight being slightly different from our assumed $R_V = 3.1$ Cardelli et al. (1989) extinction law.

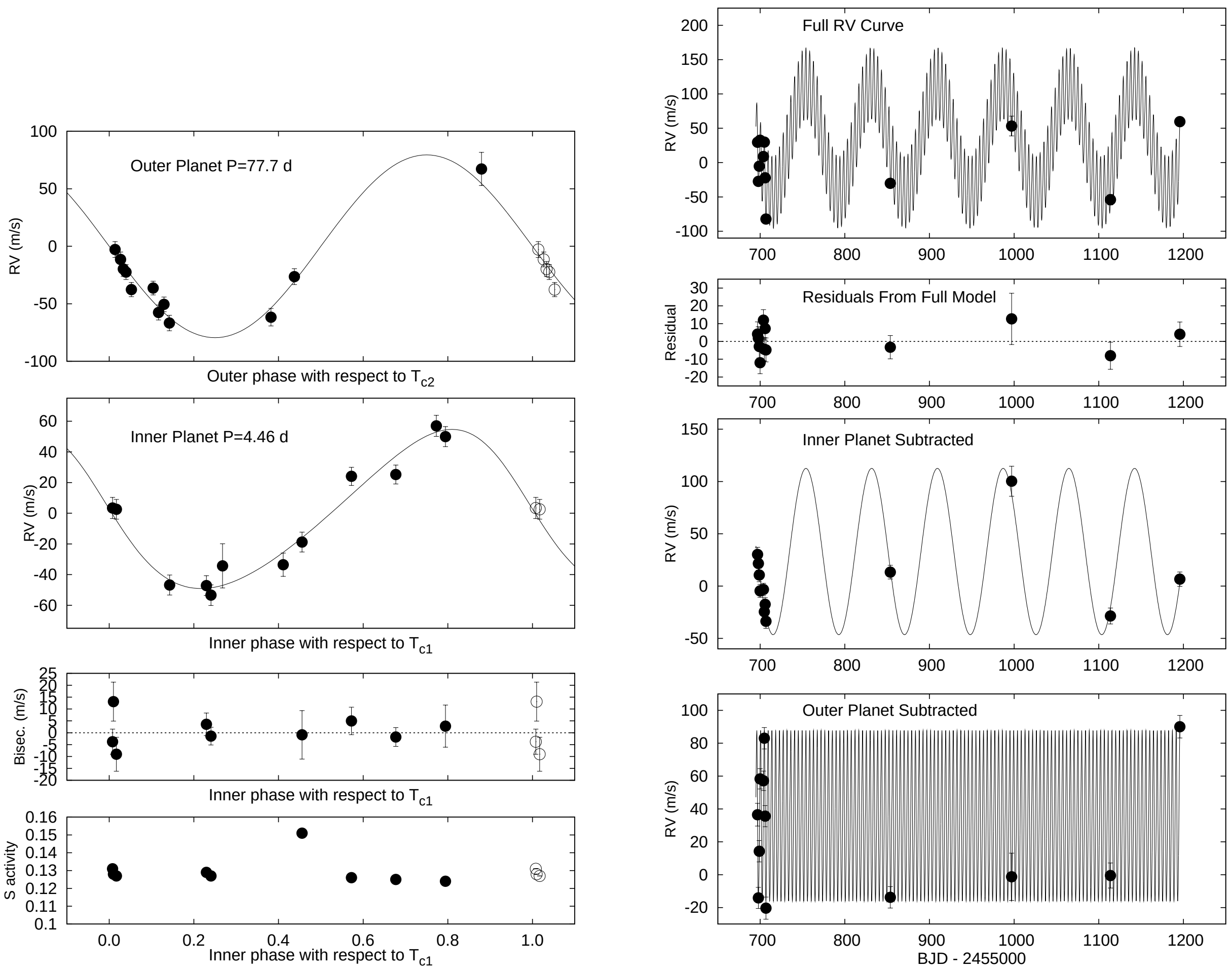

**Figure 4.** Keck/HIRES observations of HAT-P-46. The panels are as in Figure 2. The parameters used in the best-fit model are given in Table 13.

To better constrain the possible blend scenarios we obtained a partial $g$ band light curve for HAT-P-45 using Keplercam on the night of 20 May 2013. The photometry was reduced as described in Section 2.4 and included in our blend analysis procedure. We show this light curve in Figure 6, though we note that it was not included in the planet parameter determination which was carried out prior to these observations. Even though it is only a partial event, this light curve significantly restricts the range of blends that can explain the photometry for HAT-P-45, excluding scenarios that predict substantially different $g$- and $i$ band transit depths.

Although the broad-band photometry permits a wide range of possible blend scenarios, for both HAT-P-45 and HAT-P-46 the nonplanetary blend scenarios which fit the photometric data can be ruled out based on the BS and RV variations. For HAT-P-45 we find that blend scenarios that fit the photometric data (scenarios that cannot be rejected with $> 5\sigma$ confidence) yield several $\mathrm{km\,s^{-1}}$ BS and RV variations, whereas the actual BS RMS is $22\,\mathrm{m\,s^{-1}}$. Without the $g$ band light curve for HAT-P-45 some of the blend scenarios consistent with the photometry for this system predict BS and RV variations only slightly in excess of what was measured, illustrating the importance of this light curve. For HAT-P-46 the blend scenarios that fit the photometric data would result in BS variations with RMS$> 80\,\mathrm{m\,s^{-1}}$, much greater than the measured scatter of $6\,\mathrm{m\,s^{-1}}$.

We conclude that for all three objects the photometric and spectroscopic observations are best explained by transiting planets. We are not, however, able to rule out the possibility that any of these objects is actually a composite stellar system with one component hosting a transiting planet. Given the lack of definite evidence for multiple stars we analyze all of the systems assuming only one star is present in each case. If future observations identify the presence of stellar companions, the planetary masses and radii inferred in this paper will require moderate revision (e.g. Adams et al. 2013).

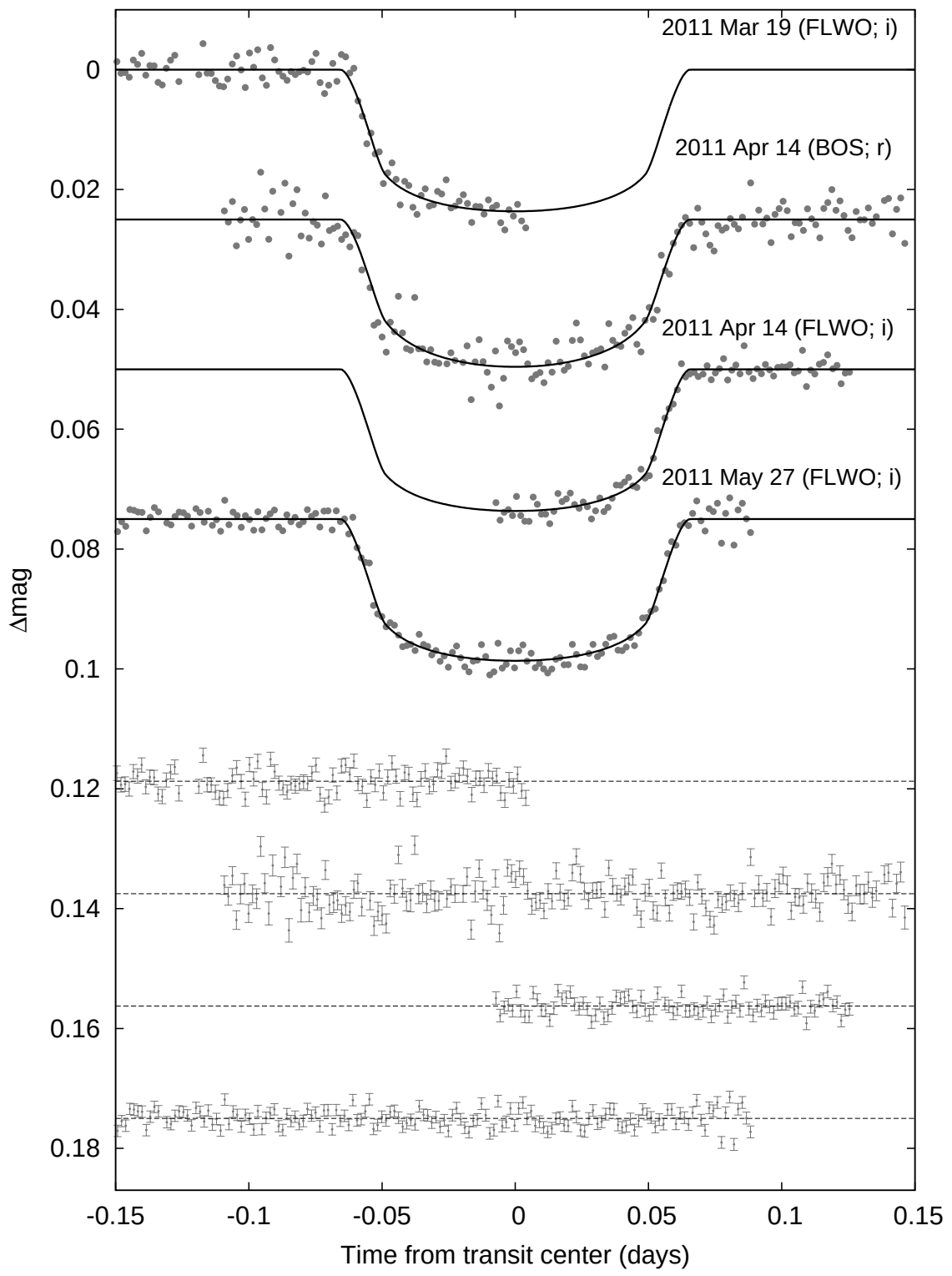

**Figure 5.** Unbinned transit light curves for HAT-P-44. The light curves have been EPD and TFA processed, as described in § 3.4. The dates of the events, the facilities used to collect the light curves, and the filters used are indicated. Curves after the first are displaced vertically for clarity. Our best fit from the global modeling described in Section 3.4 is shown by the solid lines. Residuals from the fits are displayed at the bottom, in the same order as the top curves. The error bars represent the photon and background shot noise, plus the readout noise.

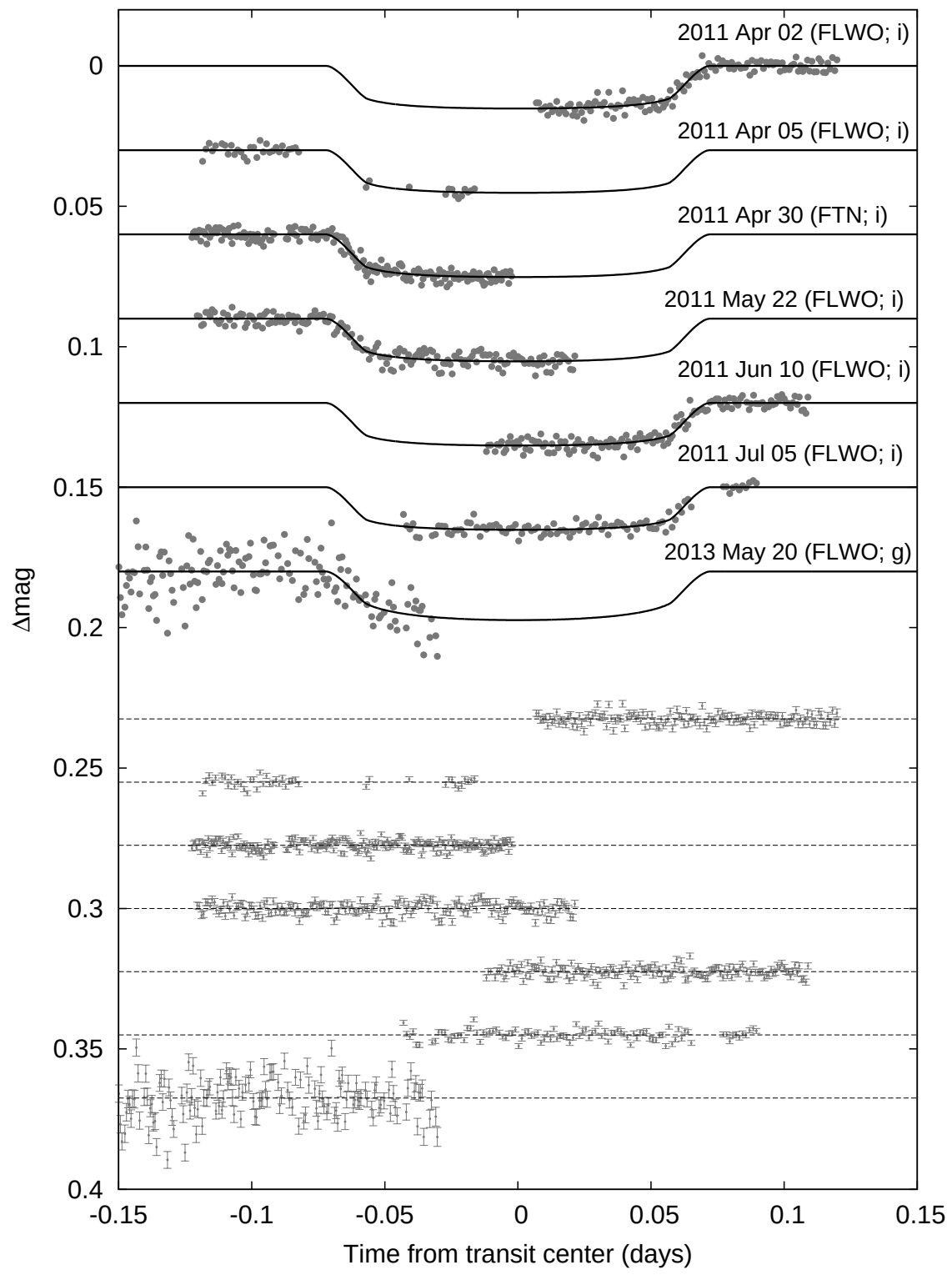

**Figure 6.** Similar to Figure 5; here we show the follow-up light curves for HAT-P-45. The $g$-band light curve obtained on 2013 May 20 was used for rejecting blend scenarios but was not used in determining the system parameters as the analysis was performed prior to carrying out this observation.

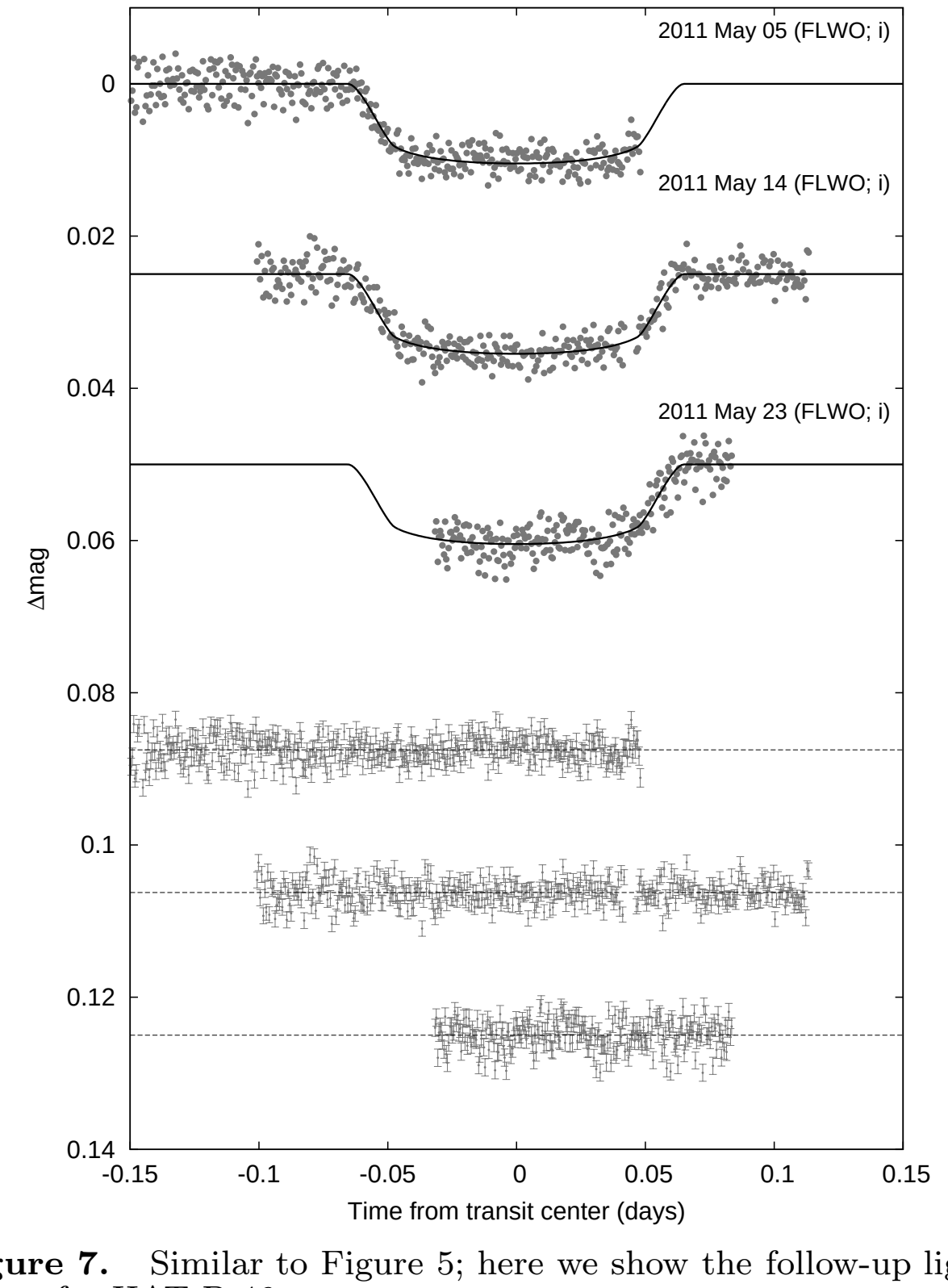

**Figure 7.** Similar to Figure 5; here we show the follow-up light curves for HAT-P-46.

### 3.3. *Periodogram Analysis of the RV Data*

For each object initial attempts to fit the data as a single planet system following the method described in Section 3.4 yielded an exceptionally high $\chi^2$ per degree of freedom (80.5, 19.7 and 23.0 for the full RV data of HAT-P-44, HAT-P-45, and HAT-P-46, respectively). Inspection of the RV residuals showed systematic variations (linear or quadratic in time) suggestive of additional components. We therefore continued to collect RV observations with Keck/HIRES for each of the objects. In all three cases the new RVs did not continue to follow the previously identified trends, indicating that if additional bodies are responsible for the excess scatter, they must have orbital periods shorter than the time-spans of the RV data sets.

Figure 9 shows the harmonic Analysis of Variance (AoV) periodograms (Schwarzenberg-Czerny 1996) of the residual RVs from the best-fit single-planet model for each system[22]. In each case strong aliasing gives rise to numerous peaks in the periodograms which could potentially phase the residual data; we are thus not able to identify a unique period for the putative outer companions in any of these systems.

[22] Using alternative methods, such as the Discrete Fourier Transform, the Lomb-Scargle periodogram, or a $\Delta\chi^2$ periodogram, yield similar frequencies, but the false alarm probabilities differ between the methods due to differences in the statistics adopted.

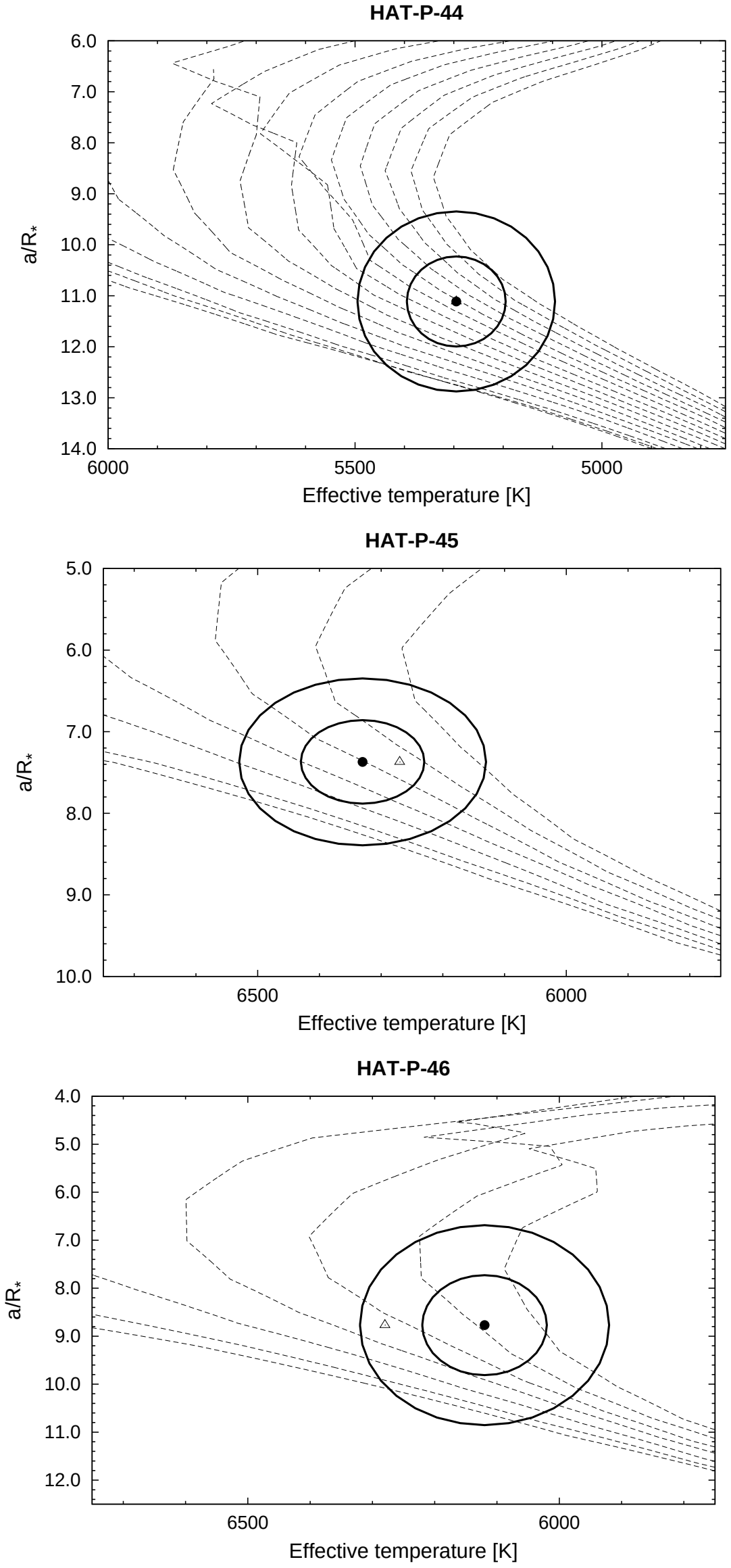

**Figure 8.** Model isochrones from Yi et al. (2001) for the metallicities of HAT-P-44 (top), HAT-P-45 (center) and HAT-P-46 (bottom). For HAT-P-44 the isochrones are shown for ages of 0.2 Gyr, 0.6 Gyr, and 1.0 Gyr to 13.0 Gyr in steps of 1.0 Gyr (left to right), while for HAT-P-44 and HAT-P-45 the isochrones are shown for ages of 0.2 Gyr, and 0.5 Gyr to 3.0 Gyr in steps of 0.5 Gyr (left to right). The adopted values of $T_{\rm eff\star}$ and $a/R_\star$ are shown together with their $1\sigma$ and $2\sigma$ confidence ellipsoids. For HAT-P-45 and HAT-P-46 the initial values of $T_{\rm eff\star}$ and $a/R_\star$ from the first SME and light curve analyses are represented with a triangle.

For HAT-P-44 the two highest peaks are at $f = 0.0023626\,{\rm d^{-1}}$ ($P = 423.26\,{\rm d}$) and $f = 0.0044477\,{\rm d^{-1}}$ ($P = 224.83\,{\rm d}$), with false alarm probabilities of $\sim 2.5\times10^{-5}$ and $3.3\times10^{-4}$, respectively. The periodogram of the residuals of a model consisting of the transiting planet and a planet with $P \sim 220\,{\rm d}$ (when fitting the data simultaneously for two planets this model provides a slightly better fit than when the outer planet has a period of $P = 423\,{\rm d}$) yields a peak at $P = 17.7\,{\rm d}$ with a false alarm probability of 0.16. Alias peaks are also seen at $P = 17.3\,{\rm d}$, $P = 18.6\,{\rm d}$, $P = 11.4\,{\rm d}$, $P = 11.7\,{\rm d}$, $P = 13.3\,{\rm d}$, and several other values with decreasing significance.

For HAT-P-45 a number of frequencies are detected in the periodogram of the RV residuals from the best-fit single-planet model. These periods are all aliases of each other. The highest peak is at $f = 0.065289\,{\rm d^{-1}}$ ($P = 15.316\,{\rm d}$), with a false alarm probability of $\sim 10^{-2}$. For HAT-P-46 the two highest peaks are at $f = 0.012977\,{\rm d^{-1}}$ ($P = 77.061\,{\rm d}$) and $f = 0.014708\,{\rm d^{-1}}$ ($P = 67.988\,{\rm d}$), each with false alarm probabilities of $\sim 10^{-2}$ (or $\sim 10^{-4}$ if uniform uncertainties are adopted as discussed further below).

The false alarm probabilities given above include a correction for the so-called "bandwidth penalty" (i.e. a correction for the number of independent frequencies that are tested by the periodogram); here we restricted the search to a frequency range of $0.02\,{\rm d^{-1}} < f < 0.2\,{\rm d^{-1}}$ and used the Horne & Baliunas (1986) approximation to estimate the number of independent frequencies tested (the resulting false alarm probability may be inaccurate by as much as a factor of $\sim 10$). Note that adopting a broader frequency range for the periodograms (e.g. up to the Nyquist limit, which for the HAT-P-44 data would be $\sim 250\,{\rm d^{-1}}$) significantly increases the false alarm probabilities. We expect, however, that systems containing multiple Jupiter-mass planets with orbital periods less than 5 days would be dynamically unstable, allowing us to restrict the frequency range to consider on physical grounds.

For HAT-P-44 and HAT-P-45 the false alarm probabilities are approximately the same for high jitter as they are when the jitter is set to 0. For HAT-P-46 the false alarm probabilities are smaller when the errors are dominated by jitter ($10^{-4}$ with jitter vs. $10^{-2}$ without jitter).

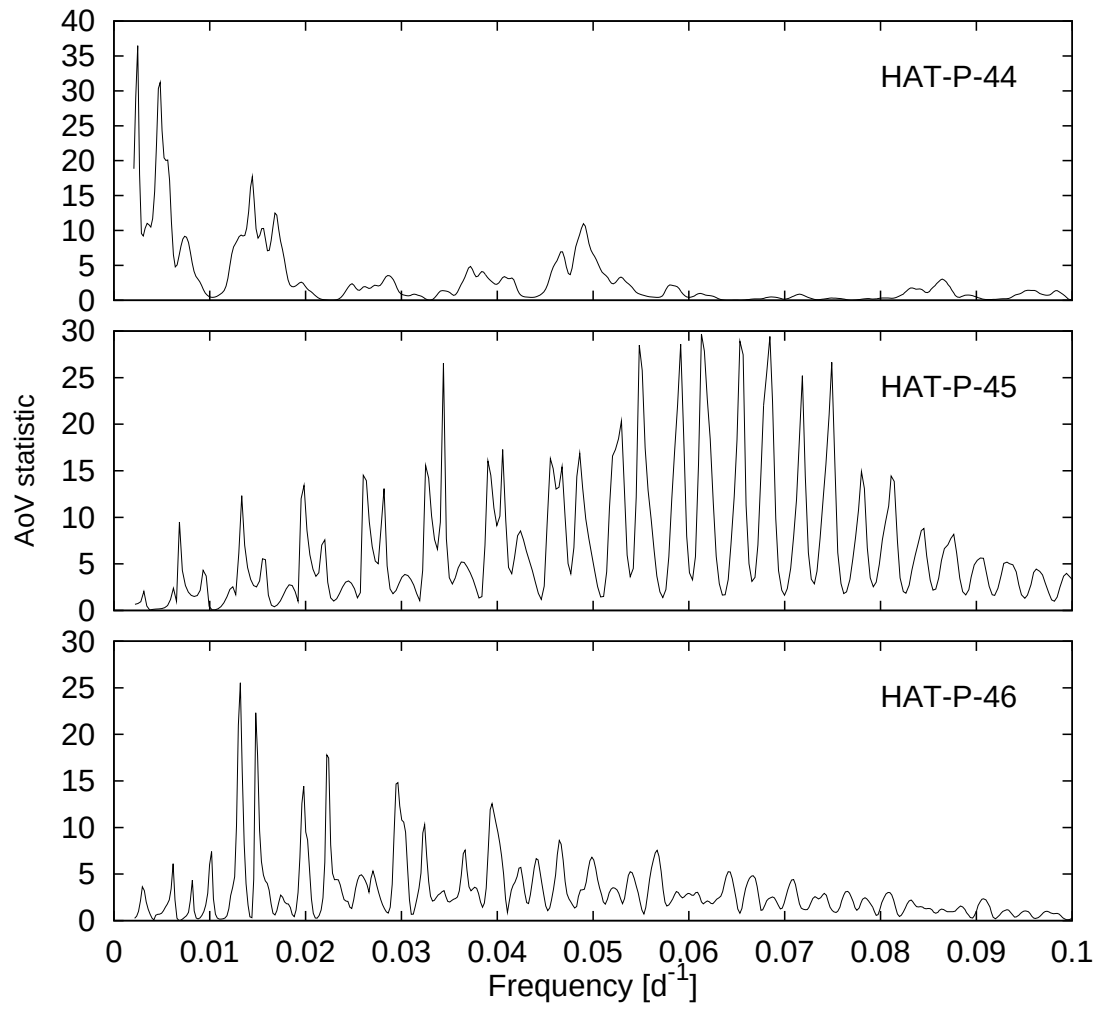

**Figure 9.** Harmonic AoV periodograms of the RV residuals from the best-fit single planet models for HAT-P-44, HAT-P-45 and HAT-P-46.

### 3.4. *Global model of the data*

We modeled simultaneously the HATNet photometry, the follow-up photometry, and the high-precision RV measurements using a procedure similar to that described in detail by Bakos et al. (2010) with modifica-tions described by Hartman et al. (2012). For each system we used a Mandel & Agol (2002) transit model, together with the EPD and TFA trend-filters, to describe the follow-up light curves, a Mandel & Agol (2002) transit model for the HATNet light curve(s), and a Keplerian orbit using the formalism of Pál (2009) for the RV curve. A significant change that we have made compared to the analysis conducted in our previous discovery papers was to include the RV jitter as a free parameter in the fit, which we discuss below. We then discuss our methods for distinguishing between competing classes of models used to fit the data, and comment on the orbital stability of potential models.

#### 3.4.1. *RV Jitter*

It is well known that high-precision RV observations of stars show non-periodic variability in excess of what is expected based on the measurement uncertainties. This "RV jitter" depends on properties of the star including the effective temperature of its photosphere, its chromospheric activity, and the projected equatorial rotation velocity of the star (see Wright 2005; Isaacson & Fischer 2010, who discuss the RV jitter from Keck/HIRES measurements). In most exoplanet studies the typical method for handling this jitter has been to add it in quadrature to the measurement uncertainties, assuming that the jitter is Gaussian white-noise. One then either adopts a jitter value that is found to be typical for similar stars, or chooses a jitter such that $\chi^2$ per degree of freedom is unity for the best-fit model. In our previous discovery papers we adopted the latter approach.

When testing competing models for the RV data the jitter is an important parameter–the greater the jitter the smaller the absolute $\chi^2$ difference between two models, and the less certain one can be in choosing one over the other. Both of the typical approaches for handling the jitter have shortcomings: the former does not allow for the possibility that a star may have a somewhat higher (or lower) than usual jitter, while the latter ignores any prior information that may be used to disfavor jitter values that would be very unusual. An alternative approach is to treat the jitter as a free parameter in the fit, but use the empirical distribution of jitters as a prior constraint.

The method of allowing the jitter to vary in an MCMC analysis of an RV curve was previously adopted by Gregory (2005). As was noted in that work, when allowing terms which appear in the uncertainties to vary in an MCMC fit, the logarithm of the likelihood is no longer simply $\ln L = -\chi^2/2 + C$ where $C$ is a normalization constant that is independent of the parameters, and can be ignored for most applications. Instead one should use $\ln L = -\chi^2/2 + \sum_{i=1}^{N} \ln(1/e_i) + C$, where $e_i$ is the error for measurement $i$ and in this case is given by $e_i = \sqrt{\sigma_i^2 + \sigma_{\rm jitter}^2}$ for formal uncertainty $\sigma_i$ and jitter $\sigma_{\rm jitter}$. When the uncertainties do not include free parameters, the term $\sum_{i=1}^{N} \ln(1/e_i)$ is constant, and included in $C$.

The analysis by Gregory (2005) used an uninformative prior on the jitter, which effectively forces the jitter to the value that results in $\chi^2/{\rm dof} = 1$; here we make use of the empirical jitter distribution found by Wright (2005) to set a prior on the jitter. Wright (2005) provides the distributions for stars in a several bins separated by $B-V$, activity, and luminosity above the main sequence. The histograms appear to be well-matched by log-normal distributions of the form:

$$P(\sigma_{\rm jitter})d\sigma_{\rm jitter} = \frac{1}{\sigma_{\rm jitter}\sqrt{2\pi\bar{\sigma}^2}} e^{-\frac{(\ln \sigma_{\rm jitter} - \bar{\mu})^2}{2\bar{\sigma}^2}} d\sigma_{\rm jitter} \tag{1}$$

Figure 10 compares this model to the jitter histograms. For HAT-P-44, which falls in the low-activity bin with $\Delta M_v < 1$ and $0.6 < B-V < 1.4$, we find $\bar{\sigma} = 0.496$, $\bar{\mu} = 1.251$, with $\sigma_{\rm jitter}$ measured in units of $\rm m\,s^{-1}$. For HAT-P-45 and HAT-P-46, which fall in the bin of low-activity stars with $\Delta M_V < 1$ and $B-V < 0.6$, we find $\bar{\sigma} = 0.688$, and $\bar{\mu} = 1.419$.

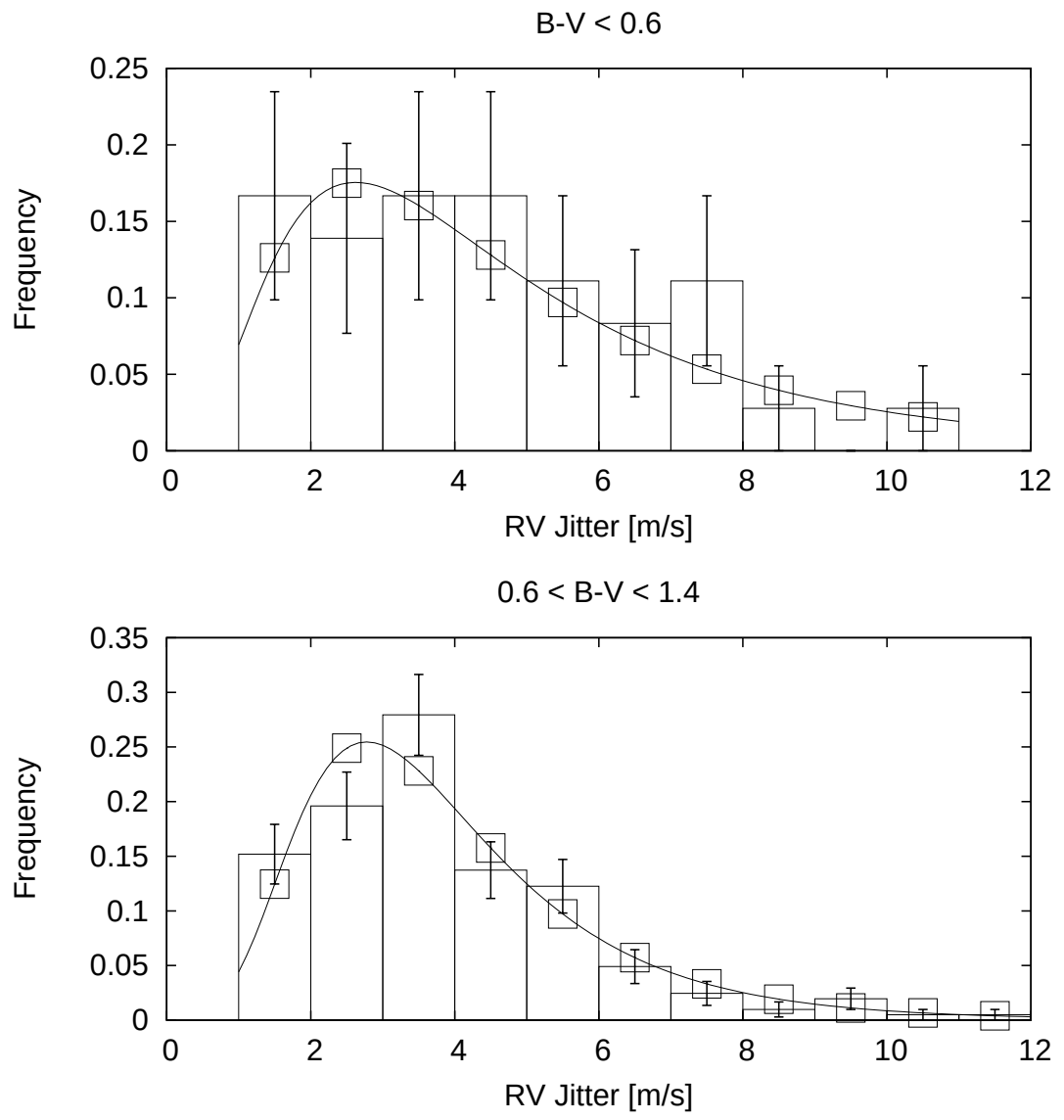

**Figure 10.** Comparison of the distribution of Keck/HIRES jitter values for low activity main sequence stars taken from Wright (2005) to our model distributions. The histograms show the empirical distributions with Poisson error-bars, the open squares show equation 1 integrated over each bin, and the solid curves are included to show the form of the distribution. The top panel is for stars with $B-V < 0.6$, appropriate for HAT-P-45 and HAT-P-46, the bottom panel is for stars with $0.6 < B-V < 1.4$, appropriate for HAT-P-44.

The posterior probability density for the parameters $\theta$, given the data $D$ and model $M$ is given by Bayes' relation:

$$P(\theta|D,M) = \frac{P(\theta|M)P(D|\theta,M)}{P(D|M)} \tag{2}$$

which in our case takes the form:

$$\begin{aligned} P(\theta|D,M) &= C\exp(\ln L + \ln P(\theta|M)) \\ &= C\exp(-\chi^2/2 + \sum_{i=1}^{N} \ln(1/e_i) + \ln P(\sigma_{\rm jitter})) \end{aligned}$$

where $C$ represents constants that are independent of $\theta$ (note we adopt uniform priors on all jump parameters other than $\sigma_{\rm jitter}$). We use a differential evolution MCMC procedure (ter Braak 2006; Eastman et al. 2013) to explore this distribution.

#### 3.4.2. *Model Selection*

As discussed in the previous subsection, modeling these objects as single-planet systems yields RV residuals with large scatter and evidence of long-term variations. We therefore performed the analysis of each system including additional Keplerian components in their RV models.

We use the Bayes Factor to select between these competing models; here we describe how this is computed. The Bayesian evidence $Z$ is defined by

$$Z = P(D|M) = \int d\theta P(\theta|M)P(D|\theta, M) \qquad (3)$$

where $P(D|M)$ is the probability of observing the data $D$ given the model $M$, marginalized over the model parameters $\theta$. The Bayes factor $K_{1,2}$ comparing the posterior probabilities for models $M_1$ and $M_2$ given the data $D$ is defined by

$$K_{1,2} = \frac{P(M_1|D)}{P(M_2|D)} = \frac{P(M_1)P(D|M_1)}{P(M_2)P(D|M_2)} \qquad (4)$$

where $P(M)$ is the prior probability for model $M$. Assuming equal priors for the different models tested, the Bayes factor is then equal to the evidence ratio:

$$K_{1,2} = \frac{Z_1}{Z_2} \qquad (5)$$

If $K_{1,2} > 1$ then model $M_1$ is favored over model $M_2$.

In practice $Z$ is difficult to determine as it requires integrating a complicated function over a high-dimensional space (e.g. Feroz et al. 2009). Recently, however, Weinberg et al. (2013) have suggested a simple and relatively accurate method for estimating $Z$ directly from the results of an MCMC simulation. Their method involves using the MCMC results to identify a small region of parameter space with high posterior probability, numerically integrating over this region, and applying a correction to scale the integral from the subregion to the full parameter space. The correction is determined from the posterior parameter distribution estimated as well from the MCMC. We use this method to estimate $Z$ and $K$ for each model. However, for practical reasons we use the MCMC sample itself to conduct a Monte Carlo integration of the parameter subregion, rather than following the suggested method of using a uniform resampling of the subregion. As shown by Weinberg et al. (2013) the method that we follow provides a somewhat biased estimate of $Z$, with errors in $\ln Z \lesssim 0.5$. For this reason we do not consider $\ln Z$ differences between models that are $< 1$ to be significant.

In Table 9 we list the models fit for each system, and provide estimates of the Bayes Factors for each model relative to a fiducial model of a single transiting planet on an eccentric orbit. For reference we also provide the Bayesian Information Criterion (BIC) estimator for each model, which is given by

$$\mathrm{BIC} = -2\ln L_{\mathrm{max}} + N_{\mathrm{p}} \ln N_{\mathrm{d}} \qquad (6)$$

for a model with $N_{\mathrm{p}}$ free parameters fit to $N_{\mathrm{d}}$ data points yielding a maximum likelihood of $L_{\mathrm{max}}$. The BIC is determined solely from the highest likelihood value, making it easier to calculate than $K$. Models with lower BIC values are generally favored. Note, however, that the BIC is a less accurate method for distinguishing between models than is $K$. We also provide, for reference, the Bayes Factors determined when the jitter of each system is fixed to a typical value throughout the analysis.

### 3.5. *Resulting Parameters*

The planet and stellar parameters for each system that are independent of the models that we test are listed in Table 10. Stellar parameters for HAT-P-44, and parameters of the transiting planet HAT-P-44b, that depend on the class of model tested are listed in Table 11, while parameters for the candidate outer components HAT-P-44c and HAT-P-44d are listed in Table 12. Stellar parameters for HAT-P-45 and HAT-P-46, and parameters for the transiting planets HAT-P-45b and HAT-P-46b, that depend on the class of model tested are listed in Table 13, while parameters for the candidate outer components HAT-P-45c, HAT-P-46c and HAT-P-46d are listed in Table 14.

For HAT-P-44 we find that the preferred model, based on the estimated Bayes Factor, consists of 2 planets, the outer one on a circular orbit. This model, labeled number 2 in Tables 9, 11, and 12, includes: the transiting planet HAT-P-44b with a period of $P = 4.301217 \pm 0.000018\,\mathrm{d}$, a mass of $M_p = 0.392 \pm 0.031\,M_{\mathrm{J}}$, and an eccentricity of $e = 0.072 \pm 0.071$; an outer planet HAT-P-44c with a period of $P = 219.9 \pm 4.5\,\mathrm{d}$, and minimum mass of $M_p \sin i = 1.6 \pm 0.2\,M_{\mathrm{J}}$. We find that an alternative model, labelled number 3, which has the same form as the preferred model, except the outer planet has a period of $P = 437.5 \pm 17.7\,\mathrm{d}$, and a minimum mass of $M_p \sin i = 3.7 \pm 0.5$ is equally acceptable based on the Bayes Factor. Note that due to the sharpness of the peaks in the likelihood as a function of the period of the outer planet, an MCMC simulation takes an excessively long time to transition between the two periods. For this reason we treat these as independent models. We adopt the model with the shorter period for the outer component because it gives a slightly higher maximum likelihood. This model is favored over the fiducial model of a single planet transiting the host star by a factor of $\sim 3 \times 10^4$ indicating that the data strongly favor the two-planet model over the single-planet model. The preferred model has an associated jitter of $12.7 \pm 2.5\,\mathrm{m\,s^{-1}}$ and a $\chi^2$ per degree of freedom, including this jitter, of 1.5. Based on equation 1, one expects only 0.5% of stars like HAT-P-44 to have jitter values $\geq 12.7 \pm 2.5\,\mathrm{m\,s^{-1}}$ thus the excess scatter in the RV residuals from the best-fit 2-planet model suggests that perhaps more than 2 planets are present in this system, though we cannot conclusively detect any additional planets from the data currently available.

For HAT-P-45 the fiducial model of a single planet on an eccentric orbit is preferred over the other models that we tested. This model, labeled number 1 under the HAT-P-45 headings in Tables 9, 13, and 14, includes only the transiting planet HAT-P-45b with a period of $P = 3.128992 \pm 0.000021\,\mathrm{d}$, a mass of $M_p = 0.892^{+0.137}_{-0.099}\,M_{\mathrm{J}}$, and an eccentricity of $e = 0.049 \pm 0.063$. The preferred model has a jitter of $22.5 \pm 6.3\,\mathrm{m\,s^{-1}}$ and $\chi^2$ per degree of freedom of 2.1. Only $\sim 0.8\%$ of stars like HAT-P-45 are expected to have a jitter this high. Moreover, the RV residuals from the preferred best-fit model appear to show a variation that is correlated in time (see the third panel down in Fig. 3). Both these factors suggest that a second planet may be present in the HAT-P-45 system. Nonetheless the data do not at present support such a complicated model. The single-planet model has a Bayes factor of $\sim 20$ relative to the two-planet model, indicating a slight preference for the single-planet model.

For HAT-P-46 the preferred model consists of a transiting planet together with an outer companion on a circular orbit. This model, labeled number 2 under the HAT-P-46 headings in Tables 9, 13 and 14, includes: the transiting planet HAT-P-46b with a period of $P = 4.463129 \pm 0.000048\,\mathrm{d}$, a mass of $M_p = 0.493^{+0.082}_{-0.052}\,M_{\mathrm{J}}$, and an eccentricity of $e = 0.123 \pm 0.120$; and an outer planet HAT-P-46c with a period of $P = 77.7 \pm 0.6\,\mathrm{d}$, and a minimum mass of $M_p \sin i = 2.0 \pm 0.3\,M_{\mathrm{J}}$. Although the two-planet model is preferred, it has a Bayes factor of only $K \sim 16$ relative to the fiducial single-planet model, indicating that the preference is not very strong. The preferred model has a jitter of $6.6 \pm 3.0\,\mathrm{m\,s^{-1}}$ and $\chi^2$ per degree of freedom of 1.6. The resulting jitter is typical for a star like HAT-P-46 ($\sim 25\%$ of such stars have a jitter higher than $6.6 \pm 3.0\,\mathrm{m\,s^{-1}}$), so there is no compelling reason at present to suspect that there may be more planets in this system beyond HAT-P-46c.

For both HAT-P-44 and HAT-P-46 allowing the jitter to vary in the fit substantially reduces the significance of the multi-planet solutions relative to the single planet solution. If we had not allowed the jitter to vary, we would have concluded that the two-planet model is $\sim 10^{39}$ times more likely than the one-planet model for HAT-P-44, and $\sim 10^{24}$ times more likely for HAT-P-46. For HAT-P-45, it is interesting to note that allowing the jitter to vary actually increases the significance of the two-planet model, perhaps due to the relatively high jitter value that must be adopted to achieve $\chi^2/\mathrm{dof} = 1$.

#### 3.5.1. *Orbital Stability*

To check the orbital stability of the multi-planet solutions that we have found, we integrated each orbital configuration forward in time for a duration of 1 Myr using the MERCURY symplectic integrator (Chambers 1999). We find that the adopted solutions for HAT-P-44 and HAT-P-46 (model 2 in each case) are stable over at least this time period, and should be stable for much longer given the large, and non-resonant, period ratio between the components in each case. For HAT-P-44 the three-planet models that we tested quickly evolved in less than $10^4$ years to a different orbital configuration. In particular, when we start HAT-P-44b on a 4.3012 d period, HAT-P-44c on a $215.7 \pm 3.9\,\mathrm{d}$ period, and HAT-P-44d on a $17.6 \pm 0.2\,\mathrm{d}$ period, HAT-P-44d migrates to a 15.1 d period orbit, while HAT-P-44b migrates to a 4.6928 d period. While this final configuration appears to be stable for at least $4 \times 10^5\,\mathrm{yr}$, it is inconsistent with the RV and photometric data. We did not carry out a full exploration of the parameter space allowed by our uncertainties, but the fact that the best-fit 3-planet model for HAT-P-44 shows rapid planetary migration indicates that this model may very well be unstable. If additional RV observations support a 3-planet solution for HAT-P-44, it will also be important to test the stability of this solution.

## 4. DISCUSSION

We have presented the discovery of three new transiting planet systems. The inner transiting planets have masses, radii, and orbital periods typical of other hot Jupiters. The planets are located on well occupied areas of both the mass–radius and the equilibrium temperature–radius diagrams. Nonetheless, as objects with well measured masses and radii, these planets will be important contributors to statistical studies of exoplanetary systems.

A notable feature of all three systems is the systematic variation seen in each of their residual RV curves. We allow in our modelling for the possibility that this excess scatter can be attributed to jitter using the empirical jitter distribution from Keck/HIRES as a prior constraint. To our knowledge this is the first time an empirical constraint on the jitter has been used in modeling the RV data for a transiting exoplanet system. Using the empirical jitter distribution significantly affects the conclusions: if we had fixed the jitter to a typical value, or a value where $\chi^2/\mathrm{dof} = 1$ for the best-fit model, we would have claimed with much greater confidence the existence of multiple planets in each system. Accounting for the uncertainty in the jitter, which must be inferred from the observations, leads to a lower confidence that we believe is more realistic.

We find that for two of the targets, HAT-P-44 and HAT-P-46, a two-planet model best explains the observations. HAT-P-44 appears to have, in addition to the $P = 4.3012\,\mathrm{d}$ transiting planet, a long period planet on a $P = 220$ or $P = 438\,\mathrm{d}$ orbit, where the ambiguity is due to aliasing. HAT-P-46 appears to have a $P = 4.4631\,\mathrm{d}$ transiting planet, and a long period planet on a $P = 78\,\mathrm{d}$ orbit, though we caution that the preference for this model over a single-planet model is not very strong for this system. Due to the limited number of RV observations, we are unable to confirm that the variation in the HAT-P-45 residual RV curve is due to a second planet, rather than being the result of anomalously high jitter for this star. Nonetheless, the high scatter, and apparent temporal correlation in that scatter, are both suggestive of a second planet. For HAT-P-44 the residuals from the best-fit two-planet model also exhibit systematic variations and scatter that is higher than expected. This may indicate the presence of a third planet in this system, however additional RV observations are needed to confirm this hypothesis.

As noted in the introduction, outer planetary companions have been confirmed for only five hot Jupiter systems ($\upsilon$ And; HD 217107; HD 187123; HIP 14810; and HAT-P-13). Only one of these, HAT-P-13, is a transiting planet system. In several other cases long-term trends have been detected, but so far the periods have not been constrained. For both HAT-P-44 and HAT-P-46 the periods of the outer planets in our adopted models are significantly longer than the transiting planet periods. This is in line with the five previously known multi-planet hot-Jupiter-bearing systems, where the shortest period outer component is HIP 14810c with a period of 147.8 d, and is unlike other multi-planet systems where densely packed systems with components having similar periods appear to be common (e.g. Lissauer et al. 2011; Fabrycky et al. 2012).

Another interesting aspect of the three systems presented here is that they all have super-solar metallicities

**Table 9**
Bayes Factor and BIC differences for models tested

| Model[a] Number | $P_c$[b] | $e_c$[c] | $P_d$ | $e_d$ | Trend Order | $N_p$[d] | ln($K$)[e] | Fixed Jitter ln($K$)[f] | ΔBIC[g] |
|---|---|---|---|---|---|---|---|---|---|
| **HAT-P-44** | | | | | | | | | |
| 2 | 219 | ··· | ··· | ··· | ··· | 7 | 10.4 | 90.7 | 46.9 |
| 3 | 426 | ··· | ··· | ··· | ··· | 7 | 10.2 | 89.7 | 45.2 |
| 4 | 221 | ✓ | ··· | ··· | ··· | 9 | 8.9 | 90.6 | 46.3 |
| 5 | 215 | ··· | 17.4 | ··· | ··· | 10 | 3.1 | 86.0 | 57.7 |
| 6 | 217 | ✓ | 17.4 | ··· | ··· | 12 | −22.5 | 82.9 | 54.9 |
| 7 | 215 | ✓ | 17.4 | ✓ | ··· | 14 | 2.5 | 82.0 | 59.2 |
| **HAT-P-45** | | | | | | | | | |
| 2 | 15.3 | ··· | ··· | ··· | ··· | 7 | −3.0 | −10.1 | 29.2 |
| 3 | ··· | ··· | ··· | ··· | 2 | 6 | −10.7 | −10.6 | −1.0 |
| **HAT-P-46** | | | | | | | | | |
| 2 | 78 | ··· | ··· | ··· | ··· | 7 | 2.8 | 55.3 | 38.8 |
| 3 | 78 | ✓ | ··· | ··· | ··· | 9 | −4.2 | 51.3 | 35.9 |
| 4 | 78 | ··· | 8.1 | ··· | ··· | 10 | −7.5 | 48.1 | 46.2 |

[a] The number associated with this model in tables 11–14. Model 1 for each system is the fiducial model of a single planet on an eccentric orbit. By definition this model has ln($K$) = 0 and ΔBIC = 0.
[b] The orbital period used for component $c$ or $d$ in days. Models for which the period of a component is listed as "···" did not include that planet.
[c] Flag indicating whether or not the component is allowed to be eccentric (indicated by a ✓), or if the eccentricity was fixed to 0 (indicated by "···").
[d] Number of varied parameters constrained by the RV observations, including 4 parameters for the inner transiting planet and one parameter for the jitter. Although the two parameters used to describe the ephemeris of the inner planet are varied in the joint fit of the RV and photometric data, they are almost entirely determined by the photometric data alone, so we do not include them in this accounting.
[e] The natural logarithm of the Bayes Factor between the given model, and a fiducial model of a single planet on an eccentric orbit. Models with higher values of $\ln K$ are preferred. In this case the RV jitter is allowed to vary in the fit, subject to a prior constraint from the empirical jitter distribution found by Wright (2005).
[f] The natural logarithm of the Bayes Factor between the given model, and a fiducial model of a single planet on an eccentric orbit. In this case the RV jitter is fixed to a typical value for each star (these were determined such that $\chi^2$ per degree of freedom was unity for one of the models; we adopted 9.1 m s$^{-1}$ for HAT-P-44, 12.7 m s$^{-1}$ for HAT-P-45, and 2.7 m s$^{-1}$ for HAT-P-46). We provide these to show how the model selection depends on the method for treating the RV jitter.
[g] $\Delta\mathrm{BIC} = \mathrm{BIC_{fiducial}} - \mathrm{BIC_{model}}$, i.e. the difference between the BIC for the fiducial model and for the given model. Models with higher values of ΔBIC are preferred.

([Fe/H]$= 0.33 \pm 0.10$, $0.07 \pm 0.10$, and $0.30 \pm 0.10$ for HAT-P-44, HAT-P-45, and HAT-P-46, respectively), as do the five previously confirmed multi-planet hot Jupiter systems ([Fe/H]$= 0.153 \pm 0.03$, $0.389 \pm 0.03$, $0.121 \pm 0.03$, $0.230 \pm 0.03$, and $0.410 \pm 0.08$ for $\upsilon$ And, HD 217107, HD 187123, HIP 14810, and HAT-P-13, respectively). That giant planets are more common around metal-rich stars is well known (Fischer & Valenti 2005); moreover, evidence suggests that the relation between metallicity and occurrence is even stronger for multi-planet systems than it is for single planet systems (e.g. Wright et al. 2009). Recently Dawson & Murray-Clay (2013) suggested that giant planets orbiting metal-rich stars are more likely to show signatures of planet-planet interactions. Of the 166 Hot-Jupiter-hosting stars in the exoplanets orbit database with measured metallicities, 109 have [Fe/H]$> 0$ and 57 have [Fe/H]$< 0$. While there is a 12% probability of finding 5 systems with [Fe/H]$> 0$ if $109/(57 + 109) = 66\%$ of systems have such a metallicity, if we include HAT-P-44 and HAT-P-46 then the probability decreases to 5%. A Kolmogorov-Smirnov (K-S) test yields a 0.6% chance that the sample of 7 multi-planet hot Jupiter systems (including HAT-P-44 and HAT-P-46) have metallicities drawn from the same distribution as all hot Jupiter-hosting systems. However, if we compare the multi-planet hot Jupiter-hosting systems to the 48 multi-planet systems with metallicities in the database that have at least one component with $M_P \sin i > 0.1\,M_J$, the K-S test yields a 23% chance that the metallicities are drawn from the same distribution. We conclude that multi-planet systems with Hot Jupiters may be more common around metal rich stars than single Hot Jupiters, to a similar extent that multi-planet systems with giant planets are in general more likely to be found around metal rich stars. A more definitive conclusion requires a careful consideration of selection effects, and a uniform determination of metallicities.

Multi-planet systems with transiting components are potentially useful for a number of applications. For example, RV observations during transit can be used to determine the projected obliquity of the transiting planet via the Rossiter-McLaughlin effect (e.g. Queloz et al. 2000). Measuring this angle for several systems would test whether the orientations of hot Jupiters in multi-planet systems are significantly different from isolated hot Jupiters, thereby testing if these two classes of systems have experienced different formation and/or evolution processes. Another example is the tidal Love number, which carries information about the interiors of planets, and can potentially be determined for transiting planets in multi-planet systems (Batygin et al. 2009; Mardling 2010; Kramm et al. 2012).

**Table 10**
Model-independent stellar and light curve parameters for HAT-P-44–HAT-P-46

| Parameter | **HAT-P-44b** Value | **HAT-P-45b** Value | **HAT-P-46b** Value | Source[a] |
|---|---|---|---|---|
| Stellar Astrometric properties | | | | |
| GSC ID | GSC 3465-00123 | GSC 5102-00262 | GSC 5100-00045 | |
| 2MASS ID | 2MASS 14123457+4700528 | 2MASS 18172957-0322517 | 2MASS 18014660-0258154 | |
| R.A. (J2000) | $14^{h}12^{m}34.56s$ | $18^{h}17^{m}29.40s$ | $18^{h}01^{m}46.56s$ | |
| Dec. (J2000) | $+47°00'52.9''$ | $-03°22'51.7''$ | $-02°58'15.4''$ | |
| $\mu_{RA}$ (mas yr$^{-1}$) | $-29.0 \pm 12.6$ | $9.2 \pm 4.1$ | $-7.5 \pm 4.3$ | |
| $\mu_{Dec}$ (mas yr$^{-1}$) | $12.3 \pm 11.4$ | $-3.1 \pm 8.2$ | $2.0 \pm 7.9$ | |
| Stellar Spectroscopic properties | | | | |
| $T_{eff\star}$ (K) | $5295 \pm 100$ | $6330 \pm 100$ | $6120 \pm 100$ | SME[b] |
| [Fe/H] | $0.33 \pm 0.10$ | $0.07 \pm 0.10$ | $0.30 \pm 0.10$ | SME |
| $v \sin i$ (km s$^{-1}$) | $0.2 \pm 0.5$ | $9.3 \pm 0.5$ | $4.9 \pm 0.5$ | SME |
| $v_{mac}$ (km s$^{-1}$) | 3.28 | 4.88 | 4.55 | SME |
| $v_{mic}$ (km s$^{-1}$) | 0.85 | 0.85 | 0.85 | SME |
| $\gamma_{RV}$ (km s$^{-1}$) | $-33.45 \pm 0.05$ | $23.903 \pm 0.1$ | $-20.911 \pm 0.1$ | TRES |
| $\log R'_{HK}$ | $-5.247$ | $-5.394$ | $-5.257$ | Keck/HIRES[c] |
| Stellar Photometric properties | | | | |
| $V$ (mag) | 13.212 | 12.794 | 11.936 | TASS |
| $V-I_C$ (mag) | $0.90 \pm 0.24$ | $1.11 \pm 0.19$ | $0.87 \pm 0.14$ | TASS |
| $J$ (mag) | $11.729 \pm 0.021$ | $10.730 \pm 0.027$ | $10.330 \pm 0.024$ | 2MASS |
| $H$ (mag) | $11.360 \pm 0.019$ | $10.350 \pm 0.026$ | $9.972 \pm 0.022$ | 2MASS |
| $K_s$ (mag) | $11.275 \pm 0.018$ | $10.201 \pm 0.023$ | $9.924 \pm 0.023$ | 2MASS |
| Transiting Planet Light curve parameters | | | | |
| $P$ (days) | $4.301217 \pm 0.000018$ | $3.128992 \pm 0.000021$ | $4.463129 \pm 0.000048$ | |
| $T_c$ (BJD) [d] | $2455352.83957 \pm 0.00136$ | $2455729.98612 \pm 0.00041$ | $2455701.33646 \pm 0.00047$ | |
| $T_{14}$ (days) [d] | $0.1303 \pm 0.0007$ | $0.1436 \pm 0.0013$ | $0.1291 \pm 0.0018$ | |
| $T_{12} = T_{34}$ (days) [d] | $0.0158 \pm 0.0006$ | $0.0154 \pm 0.0011$ | $0.0174 \pm 0.0017$ | |
| $\zeta/R_\star$ | $17.48 \pm 0.07$ | $15.60 \pm 0.08$ | $17.83 \pm 0.13$ | |
| $R_p/R_\star$ | $0.1343 \pm 0.0011$ | $0.1110 \pm 0.0021$ | $0.0942 \pm 0.0017$ | |
| $b^2$ | $0.030^{+0.036}_{-0.018}$ | $0.079^{+0.058}_{-0.043}$ | $0.392^{+0.047}_{-0.059}$ | |
| $b \equiv a \cos i/R_\star$ | $0.173^{+0.076}_{-0.075}$ | $0.281^{+0.085}_{-0.106}$ | $0.626^{+0.036}_{-0.051}$ | |
| Assumed Limb-darkening coefficients [e] | | | | |
| $c_1, i$ (linear term) | 0.3648 | 0.1935 | 0.2222 | Claret 2004 |
| $c_2, i$ (quadratic term) | 0.2817 | 0.3680 | 0.3651 | Claret 2004 |

[a] We list the source only for the stellar properties. The listed transiting planet light curve parameters are determined from our joint fit of the RV and light curve data, but are primarily constrained by the light curves.
[b] SME = "Spectroscopy Made Easy" package for the analysis of high-resolution spectra (Valenti & Piskunov 1996). These parameters rely primarily on SME, but have a small dependence also on the iterative analysis incorporating the isochrone search and global modeling of the data, as described in the text.
[c] Median values of $\log R'_{HK}$ (Noyes et al. 1984) are computed from the Keck/HIRES spectra following the procedure of Isaacson & Fischer (2010).
[d] $T_c$: Reference epoch of mid transit that minimizes the correlation with the orbital period. $T_{14}$: total transit duration, time between first to last contact; $T_{12} = T_{34}$: ingress/egress time, time between first and second, or third and fourth contact. Barycentric Julian dates (BJD) throughout the paper are calculated from Coordinated Universal Time (UTC).
[e] Values for a quadratic law.

We stress that each of the systems presented here would greatly benefit from continued long-term RV monitoring to confirm the outer planets and characterize their properties.

HATNet operations have been funded in part by NASA grants NNG04GN74G, NNX08AF23G and NNX13AJ15G. J.D.H., G.Á.B. and W.B. received partial support from NSF grant AST-1108686. G.Á.B., Z.C. and K.P. acknowledge partial support from NASA grant NNX09AB29G. GT acknowledges partial support from NASA grant NNX09AF59G. We acknowledge partial support also from the Kepler Mission under NASA Cooperative Agreement NCC2-1390 (D.W.L., PI). G.K. thanks the Hungarian Scientific Research Foundation (OTKA) for support through grant K-81373. This research has made use of Keck telescope time granted through NOAO (program A284Hr) and NASA (N154Hr, N108Hr). This paper uses observations obtained with facilities of the Las Cumbres Observatory Global Telescope. Data presented in this paper are based on observations obtained at the HAT station at the Submillimeter Array of SAO, and the HAT station at the Fred Lawrence Whipple Observatory of SAO. We also thank Mount Stromlo Observatory and Siding Spring Observatory for granting us time on the ANU 2.3 m telescope. This research has made use of the Exoplanet Orbit

**Table 11**
Model-dependent system parameters for HAT-P-44

| Parameter | Model 1 Value | **Adopted Model 2 Value** | Model 3 Value | Model 4 Value | Model 5 Value | Model 6 Value | Model 7 Value |
|---|---|---|---|---|---|---|---|
| Transiting planet (HAT-P-44b) light curve parameters | | | | | | | |
| $a/R_\star$ | $9.34 \pm 1.70$ | $11.11^{+0.70}_{-1.03}$ | $10.13 \pm 1.08$ | $11.37^{+0.54}_{-0.97}$ | $11.35^{+0.48}_{-0.76}$ | $11.25^{+0.12}_{-0.23}$ | $11.33^{+0.46}_{-0.70}$ |
| $i$ (deg) | $88.7^{+0.6}_{-1.6}$ | $89.0 \pm 0.5$ | $88.8^{+0.5}_{-0.8}$ | $89.1 \pm 0.5$ | $89.2 \pm 0.4$ | $89.1 \pm 0.4$ | $89.1^{+0.4}_{-0.5}$ |
| Transiting planet (HAT-P-44b) RV parameters | | | | | | | |
| $K$ ($\mathrm{m\,s^{-1}}$) | $47.1 \pm 10.7$ | $51.5 \pm 3.9$ | $52.5 \pm 4.7$ | $48.1 \pm 4.1$ | $52.6 \pm 3.4$ | $52.1 \pm 3.2$ | $53.1 \pm 3.3$ |
| $\sqrt{e}\cos\omega$ | $-0.206^{+0.257}_{-0.189}$ | $-0.006 \pm 0.121$ | $-0.113 \pm 0.113$ | $-0.046 \pm 0.125$ | $-0.040 \pm 0.107$ | $-0.026 \pm 0.106$ | $-0.034 \pm 0.105$ |
| $\sqrt{e}\sin\omega$ | $0.421^{+0.172}_{-0.312}$ | $0.219^{+0.151}_{-0.236}$ | $0.364^{+0.122}_{-0.231}$ | $0.147 \pm 0.185$ | $0.171^{+0.134}_{-0.189}$ | $0.204 \pm 0.000$ | $0.175^{+0.120}_{-0.179}$ |
| $e\cos\omega$ | $-0.099 \pm 0.129$ | $-0.001 \pm 0.037$ | $-0.041 \pm 0.047$ | $-0.008^{+0.030}_{-0.042}$ | $-0.007^{+0.024}_{-0.032}$ | $-0.005 \pm 0.029$ | $-0.006^{+0.023}_{-0.032}$ |
| $e\sin\omega$ | $0.214 \pm 0.170$ | $0.055^{+0.098}_{-0.059}$ | $0.144 \pm 0.103$ | $0.029^{+0.092}_{-0.042}$ | $0.035^{+0.071}_{-0.039}$ | $0.044^{+0.009}_{-0.002}$ | $0.036^{+0.063}_{-0.037}$ |
| $e$ | $0.272 \pm 0.155$ | $0.072 \pm 0.071$ | $0.158 \pm 0.098$ | $0.054 \pm 0.062$ | $0.050 \pm 0.050$ | $0.048 \pm 0.016$ | $0.049 \pm 0.046$ |
| $\omega$ (deg) | $117 \pm 50$ | $98 \pm 84$ | $108 \pm 44$ | $114 \pm 82$ | $107 \pm 75$ | $97 \pm 25$ | $105 \pm 72$ |
| RV jitter ($\mathrm{m\,s^{-1}}$) | $31.0 \pm 4.2$ | $12.7 \pm 2.5$ | $13.4 \pm 2.3$ | $11.5 \pm 2.1$ | $9.1 \pm 1.9$ | $8.4 \pm 2.1$ | $7.9 \pm 2.2$ |
| Derived transiting planet (HAT-P-44b) parameters | | | | | | | |
| $M_p$ ($M_{\rm J}$) | $0.347 \pm 0.077$ | $0.392 \pm 0.031$ | $0.394 \pm 0.036$ | $0.368 \pm 0.033$ | $0.403 \pm 0.028$ | $0.397 \pm 0.027$ | $0.406 \pm 0.027$ |
| $R_p$ ($R_{\rm J}$) | $1.523^{+0.442}_{-0.226}$ | $1.280^{+0.145}_{-0.074}$ | $1.403^{+0.190}_{-0.130}$ | $1.256^{+0.126}_{-0.059}$ | $1.255^{+0.095}_{-0.052}$ | $1.263^{+0.036}_{-0.026}$ | $1.255^{+0.090}_{-0.051}$ |
| $C(M_p, R_p)$ [a] | 0.18 | 0.10 | 0.11 | 0.16 | 0.10 | 0.18 | 0.18 |
| $\rho_p$ ($\mathrm{g\,cm^{-3}}$) | $0.12^{+0.09}_{-0.05}$ | $0.23 \pm 0.05$ | $0.18^{+0.07}_{-0.05}$ | $0.23 \pm 0.05$ | $0.25 \pm 0.04$ | $0.24 \pm 0.02$ | $0.25 \pm 0.04$ |
| $\log g_p$ (cgs) | $2.55 \pm 0.19$ | $2.77^{+0.06}_{-0.09}$ | $2.69 \pm 0.10$ | $2.75^{+0.06}_{-0.08}$ | $2.80^{+0.05}_{-0.07}$ | $2.79^{+0.03}_{-0.04}$ | $2.80^{+0.04}_{-0.06}$ |
| $a$ (AU) | $0.0509^{+0.0014}_{-0.0008}$ | $0.0507 \pm 0.0007$ | $0.0507 \pm 0.0008$ | $0.0507 \pm 0.0007$ | $0.0507 \pm 0.0007$ | $0.0506 \pm 0.0007$ | $0.0507 \pm 0.0007$ |
| $T_{\rm eq}$ (K) | $1238^{+173}_{-107}$ | $1126^{+67}_{-42}$ | $1181^{+84}_{-64}$ | $1114^{+59}_{-36}$ | $1114^{+47}_{-33}$ | $1118 \pm 23$ | $1114^{+44}_{-32}$ |
| $\Theta$ [b] | $0.024 \pm 0.007$ | $0.033 \pm 0.003$ | $0.030 \pm 0.004$ | $0.031 \pm 0.003$ | $0.034 \pm 0.003$ | $0.034 \pm 0.002$ | $0.035 \pm 0.003$ |
| $\langle F \rangle$ [c] | $5.30^{+4.25}_{-1.54}$ | $3.63^{+1.02}_{-0.50}$ | $4.40^{+1.49}_{-0.85}$ | $3.48^{+0.86}_{-0.42}$ | $3.48^{+0.66}_{-0.38}$ | $3.53 \pm 0.29$ | $3.48^{+0.62}_{-0.37}$ |
| Derived stellar properties | | | | | | | |
| $M_\star$ ($M_\odot$) | $0.953^{+0.083}_{-0.045}$ | $0.939 \pm 0.041$ | $0.938 \pm 0.042$ | $0.941 \pm 0.041$ | $0.940 \pm 0.041$ | $0.937 \pm 0.040$ | $0.940 \pm 0.041$ |
| $R_\star$ ($R_\odot$) | $1.165^{+0.334}_{-0.173}$ | $0.979^{+0.110}_{-0.055}$ | $1.072^{+0.144}_{-0.099}$ | $0.960^{+0.096}_{-0.043}$ | $0.961^{+0.072}_{-0.039}$ | $0.969 \pm 0.021$ | $0.961^{+0.067}_{-0.037}$ |
| $\log g_\star$ (cgs) | $4.28 \pm 0.16$ | $4.43 \pm 0.07$ | $4.35 \pm 0.09$ | $4.45 \pm 0.06$ | $4.45 \pm 0.05$ | $4.44 \pm 0.02$ | $4.44 \pm 0.05$ |
| $L_\star$ ($L_\odot$) | $0.96^{+0.71}_{-0.27}$ | $0.68^{+0.19}_{-0.10}$ | $0.81^{+0.27}_{-0.16}$ | $0.66^{+0.16}_{-0.09}$ | $0.66^{+0.13}_{-0.08}$ | $0.66 \pm 0.07$ | $0.66^{+0.12}_{-0.08}$ |
| $M_V$ (mag) | $4.97 \pm 0.46$ | $5.34 \pm 0.23$ | $5.15 \pm 0.28$ | $5.38 \pm 0.21$ | $5.38 \pm 0.18$ | $5.38 \pm 0.14$ | $5.38 \pm 0.18$ |
| $M_K$ (mag,ESO) | $3.06 \pm 0.44$ | $3.44 \pm 0.19$ | $3.24 \pm 0.24$ | $3.48 \pm 0.16$ | $3.48 \pm 0.13$ | $3.46 \pm 0.07$ | $3.47 \pm 0.13$ |
| Age (Gyr) | $11.5^{+3.3}_{-4.4}$ | $8.9 \pm 3.9$ | $11.5 \pm 3.8$ | $8.1 \pm 3.8$ | $8.1 \pm 3.5$ | $8.5 \pm 2.7$ | $8.1 \pm 3.4$ |
| $A_V$ (mag) | $0.000 \pm 0.083$ | $0.000 \pm 0.081$ | $0.000 \pm 0.083$ | $0.000 \pm 0.080$ | $0.000 \pm 0.081$ | $0.000 \pm 0.081$ | $0.000 \pm 0.082$ |
| Distance (pc) | $445^{+127}_{-66}$ | $374^{+42}_{-23}$ | $409^{+55}_{-39}$ | $367^{+37}_{-19}$ | $367^{+28}_{-17}$ | $370 \pm 11$ | $367^{+26}_{-17}$ |

[a] Correlation coefficient between the planetary mass $M_p$ and radius $R_p$.
[b] The Safronov number is given by $\Theta = \frac{1}{2}(V_{\rm esc}/V_{\rm orb})^2 = (a/R_p)(M_p/M_\star)$ (see Hansen & Barman 2007).
[c] Incoming flux per unit surface area, averaged over the orbit, measured in units of $10^8\mathrm{erg\,s^{-1}\,cm^{-2}}$.

Database and the Exoplanet Data Explorer at exoplanets.org.

**Table 12**
Model-dependent parameters for outer planets in HAT-P-44

| Parameter | Model 1 Value | **Adopted Model 2 Value** | Model 3 Value | Model 4 Value | Model 5 Value | Model 6 Value | Model 7 Value |
|---|---|---|---|---|---|---|---|
| RV and derived parameters for candidate planet HAT-P-44c | | | | | | | |
| $P_c$ (days) ..... | ··· | $219.9 \pm 4.5$ | $437.5 \pm 17.7$ | $221.4 \pm 4.4$ | $218.7 \pm 5.6$ | $218.6 \pm 5.7$ | $215.7 \pm 3.9$ |
| $T_{c,c}$ (BJD) [a] . | ··· | $2455686.2 \pm 4.1$ | $2455727.0 \pm 4.5$ | $2455713.4 \pm 4.3$ | $2455685.7 \pm 3.2$ | $2455682.1 \pm 2.9$ | $2455686.2 \pm 6.1$ |
| $T_{14,c}$ (days) [a] | ··· | $0.505 \pm 0.039$ | $0.691 \pm 0.067$ | $0.652 \pm 0.091$ | $0.496 \pm 0.026$ | $0.524 \pm 0.039$ | $0.567 \pm 0.067$ |
| $K_c$ ($\mathrm{m\,s^{-1}}$) ... | ··· | $56 \pm 6$ | $104 \pm 12$ | $93 \pm 4$ | $54 \pm 5$ | $53 \pm 3$ | $64 \pm 10$ |
| $\sqrt{e}\cos\omega_c$ ..... | ··· | 0 | 0 | $-0.373 \pm 0.105$ | 0 | $0.194^{+0.105}_{-0.227}$ | $0.090 \pm 0.145$ |
| $\sqrt{e}\sin\omega_c$ ..... | ··· | 0 | 0 | $-0.474 \pm 0.113$ | 0 | $-0.172 \pm 0.162$ | $-0.351^{+0.254}_{-0.130}$ |
| $e\cos\omega_c$ ....... | ··· | 0 | 0 | $-0.226 \pm 0.071$ | 0 | $0.055 \pm 0.066$ | $0.031 \pm 0.058$ |
| $e\sin\omega_c$ ....... | ··· | 0 | 0 | $-0.286^{+0.079}_{-0.103}$ | 0 | $-0.049^{+0.053}_{-0.083}$ | $-0.136 \pm 0.107$ |
| $e_c$ ........... | ··· | 0 | 0 | $0.379 \pm 0.075$ | 0 | $0.096 \pm 0.071$ | $0.153 \pm 0.093$ |
| $\omega_c$ (deg) ...... | ··· | 0 | 0 | $232.1 \pm 13.3$ | 0 | $305.7 \pm 104.3$ | $279.8 \pm 70.0$ |
| $M_p \sin i_c$ ($M_J$) | ··· | $1.6 \pm 0.2$ | $3.7 \pm 0.5$ | $2.5 \pm 0.1$ | $1.5 \pm 0.2$ | $1.5 \pm 0.1$ | $1.8^{+0.3}_{-0.2}$ |
| $a_c$ (AU) ...... | ··· | $0.699 \pm 0.014$ | $1.104^{+0.039}_{-0.027}$ | $0.702 \pm 0.014$ | $0.696 \pm 0.016$ | $0.695 \pm 0.016$ | $0.690 \pm 0.013$ |
| RV and derived parameters for candidate planet HAT-P-44d | | | | | | | |
| $P_d$ (days) ..... | ··· | ··· | ··· | ··· | $17.4 \pm 0.2$ | $17.4 \pm 0.3$ | $17.6 \pm 0.2$ |
| $T_{c,d}$ (BJD) [a] . | ··· | ··· | ··· | ··· | $2455699.6 \pm 0.6$ | $2455699.4 \pm 0.5$ | $2455699.5 \pm 0.7$ |
| $T_{14,d}$ (days) [a] | ··· | ··· | ··· | ··· | $0.213 \pm 0.012$ | $0.215 \pm 0.004$ | $0.185 \pm 0.065$ |
| $K_d$ ($\mathrm{m\,s^{-1}}$) ... | ··· | ··· | ··· | ··· | $16 \pm 4$ | $19 \pm 4$ | $20 \pm 8$ |
| $\sqrt{e}\cos\omega_d$ ..... | ··· | ··· | ··· | ··· | 0 | 0 | $0.071 \pm 0.368$ |
| $\sqrt{e}\sin\omega_d$ ..... | ··· | ··· | ··· | ··· | 0 | 0 | $0.292 \pm 0.371$ |
| $e\cos\omega_d$ ....... | ··· | ··· | ··· | ··· | 0 | 0 | $0.019^{+0.359}_{-0.183}$ |
| $e\sin\omega_d$ ....... | ··· | ··· | ··· | ··· | 0 | 0 | $0.128^{+0.390}_{-0.176}$ |
| $e_d$ ........... | ··· | ··· | ··· | ··· | 0 | 0 | $0.251 \pm 0.314$ |
| $\omega_d$ (deg) ...... | ··· | ··· | ··· | ··· | 0 | 0 | $102.4 \pm 98.4$ |
| $M_p \sin i_d$ ($M_J$) | ··· | ··· | ··· | ··· | $0.2 \pm 0.1$ | $0.2 \pm 0.1$ | $0.2 \pm 0.0$ |
| $a_d$ (AU) ...... | ··· | ··· | ··· | ··· | $0.129 \pm 0.002$ | $0.129 \pm 0.002$ | $0.130 \pm 0.002$ |

[a] $T_c$: Reference epoch of mid transit that minimizes the correlation with the orbital period. $T_{14}$: total transit duration, time between first to last contact; Barycentric Julian dates (BJD) throughout the paper are calculated from Coordinated Universal Time (UTC).

**Table 13**
Model-dependent system parameters for HAT-P-45 and HAT-P-46

| Parameter | HAT-P-45 **Adopted Model 1 Value** | HAT-P-45 Model 2 Value | HAT-P-45 Model 3 Value | HAT-P-46 Model 1 Value | HAT-P-46 **Adopted Model 2 Value** | HAT-P-46 Model 3 Value | HAT-P-46 Model 4 Value |
|---|---|---|---|---|---|---|---|
| Transiting planet (HAT-P-45b and HAT-P-46b) light curve parameters | | | | | | | |
| $a/R_\star$ .......... | $7.36^{+0.39}_{-0.62}$ | $7.56 \pm 0.31$ | $7.39^{+0.43}_{-0.66}$ | $8.84^{+1.04}_{-1.76}$ | $8.86^{+0.89}_{-1.24}$ | $9.27^{+0.66}_{-1.24}$ | $9.75^{+0.49}_{-0.81}$ |
| $i$ (deg) ......... | $87.8 \pm 0.9$ | $87.9 \pm 0.8$ | $87.8 \pm 0.9$ | $85.4^{+1.0}_{-4.1}$ | $85.5^{+0.8}_{-2.3}$ | $85.9^{+0.6}_{-2.2}$ | $86.3^{+0.4}_{-0.9}$ |
| Transiting planet (HAT-P-45b and HAT-P-46b) RV parameters | | | | | | | |
| $K$ ($\mathrm{m\,s^{-1}}$) ...... | $106.6 \pm 13.6$ | $106.9 \pm 4.5$ | $104.1 \pm 12.7$ | $40.9 \pm 19.7$ | $52.2 \pm 6.8$ | $48.3 \pm 5.2$ | $49.7 \pm 3.5$ |
| $\sqrt{e}\cos\omega$ ........ | $-0.009 \pm 0.160$ | $-0.031 \pm 0.086$ | $-0.070^{+0.201}_{-0.148}$ | $0.180^{+0.239}_{-0.317}$ | $0.143^{+0.098}_{-0.134}$ | $0.108 \pm 0.118$ | $0.013 \pm 0.105$ |
| $\sqrt{e}\sin\omega$ ......... | $0.045 \pm 0.192$ | $-0.084 \pm 0.131$ | $0.021 \pm 0.202$ | $0.242 \pm 0.269$ | $0.305^{+0.170}_{-0.258}$ | $0.201 \pm 0.215$ | $0.020^{+0.210}_{-0.158}$ |
| $e\cos\omega$ .......... | $-0.001^{+0.061}_{-0.045}$ | $-0.004 \pm 0.017$ | $-0.012 \pm 0.061$ | $0.064^{+0.183}_{-0.140}$ | $0.047^{+0.066}_{-0.042}$ | $0.027^{+0.062}_{-0.032}$ | $0.001^{+0.029}_{-0.021}$ |
| $e\sin\omega$ ........... | $0.006^{+0.089}_{-0.045}$ | $-0.010^{+0.024}_{-0.039}$ | $0.002^{+0.093}_{-0.054}$ | $0.091^{+0.217}_{-0.108}$ | $0.105^{+0.149}_{-0.089}$ | $0.050^{+0.152}_{-0.056}$ | $0.002^{+0.084}_{-0.036}$ |
| $e$ ............... | $0.049 \pm 0.063$ | $0.025 \pm 0.026$ | $0.061 \pm 0.066$ | $0.195 \pm 0.167$ | $0.123 \pm 0.120$ | $0.073 \pm 0.111$ | $0.031 \pm 0.054$ |
| $\omega$ (deg) ......... | $146 \pm 98$ | $239 \pm 87$ | $168 \pm 88$ | $82 \pm 113$ | $70 \pm 87$ | $75 \pm 105$ | $154 \pm 103$ |
| RV jitter ($\mathrm{m\,s^{-1}}$) | $22.5 \pm 6.3$ | $4.7 \pm 3.5$ | $22.5 \pm 6.6$ | $28.4 \pm 6.7$ | $6.6 \pm 3.0$ | $5.7 \pm 2.7$ | $3.0 \pm 2.2$ |
| $\dot{\gamma}$ ($\mathrm{m\,s^{-1}\,d^{-1}}$)[d] .. | ··· | ··· | $-0.2617 \pm 0.3900$ | ··· | ··· | ··· | ··· |
| $\ddot{\gamma}$ ($\mathrm{m\,s^{-1}\,d^{-2}}$)[d] .. | ··· | ··· | $0.0011 \pm 0.0013$ | ··· | ··· | ··· | ··· |
| Derived transiting planet (HAT-P-45b and HAT-P-46b) parameters | | | | | | | |
| $M_p$ ($M_{\rm J}$) ........ | $0.892^{+0.137}_{-0.099}$ | $0.890 \pm 0.046$ | $0.869 \pm 0.111$ | $0.383^{+0.218}_{-0.127}$ | $0.493^{+0.082}_{-0.052}$ | $0.455^{+0.063}_{-0.039}$ | $0.466 \pm 0.036$ |
| $R_p$ ($R_{\rm J}$) ........ | $1.426^{+0.175}_{-0.087}$ | $1.382 \pm 0.076$ | $1.422^{+0.184}_{-0.095}$ | $1.286^{+0.426}_{-0.150}$ | $1.284^{+0.271}_{-0.133}$ | $1.222^{+0.257}_{-0.096}$ | $1.157^{+0.142}_{-0.071}$ |
| $C(M_p, R_p)$ [a] .... | 0.45 | 0.40 | 0.33 | 0.43 | 0.72 | 0.63 | 0.41 |
| $\rho_p$ ($\mathrm{g\,cm^{-3}}$) ..... | $0.38 \pm 0.09$ | $0.42 \pm 0.06$ | $0.37 \pm 0.10$ | $0.20^{+0.14}_{-0.09}$ | $0.28 \pm 0.10$ | $0.30 \pm 0.09$ | $0.37 \pm 0.08$ |
| $\log g_p$ (cgs) ...... | $3.03 \pm 0.07$ | $3.06 \pm 0.04$ | $3.02 \pm 0.08$ | $2.72^{+0.16}_{-0.28}$ | $2.86 \pm 0.10$ | $2.87^{+0.07}_{-0.12}$ | $2.93^{+0.06}_{-0.08}$ |
| $a$ (AU) .......... | $0.0452 \pm 0.0007$ | $0.0451 \pm 0.0006$ | $0.0452 \pm 0.0007$ | $0.0577^{+0.0020}_{-0.0010}$ | $0.0577^{+0.0014}_{-0.0009}$ | $0.0574^{+0.0013}_{-0.0008}$ | $0.0570 \pm 0.0008$ |
| $T_{\rm eq}$ (K) ......... | $1652^{+90}_{-52}$ | $1627 \pm 44$ | $1649^{+95}_{-56}$ | $1465^{+220}_{-89}$ | $1458^{+140}_{-75}$ | $1425^{+135}_{-57}$ | $1386^{+74}_{-44}$ |
| $\Theta$[b] .............. | $0.044 \pm 0.006$ | $0.046 \pm 0.003$ | $0.043 \pm 0.006$ | $0.025 \pm 0.009$ | $0.034 \pm 0.004$ | $0.033 \pm 0.004$ | $0.036 \pm 0.004$ |
| $\langle F \rangle$ [c] ........... | $1.68^{+0.44}_{-0.20}$ | $1.58 \pm 0.17$ | $1.67^{+0.47}_{-0.21}$ | $1.04^{+0.89}_{-0.22}$ | $1.02^{+0.50}_{-0.19}$ | $9.30^{+4.86}_{-1.37}$ | $8.33^{+2.13}_{-0.99}$ |
| Derived stellar properties | | | | | | | |
| $M_\star$ ($M_\odot$) ....... | $1.259 \pm 0.058$ | $1.246 \pm 0.050$ | $1.257 \pm 0.059$ | $1.289^{+0.144}_{-0.068}$ | $1.284^{+0.095}_{-0.060}$ | $1.266^{+0.090}_{-0.054}$ | $1.243 \pm 0.052$ |
| $R_\star$ ($R_\odot$) ........ | $1.319^{+0.155}_{-0.072}$ | $1.281 \pm 0.060$ | $1.314^{+0.165}_{-0.079}$ | $1.398^{+0.465}_{-0.156}$ | $1.396^{+0.293}_{-0.136}$ | $1.329^{+0.279}_{-0.095}$ | $1.257^{+0.144}_{-0.067}$ |
| $\log g_\star$ (cgs)...... | $4.30 \pm 0.06$ | $4.32 \pm 0.03$ | $4.30 \pm 0.07$ | $4.25 \pm 0.15$ | $4.25 \pm 0.11$ | $4.29 \pm 0.10$ | $4.33 \pm 0.06$ |
| $L_\star$ ($L_\odot$) ........ | $2.51^{+0.71}_{-0.33}$ | $2.35 \pm 0.30$ | $2.49^{+0.75}_{-0.36}$ | $2.48^{+2.08}_{-0.57}$ | $2.46^{+1.25}_{-0.49}$ | $2.23^{+1.18}_{-0.36}$ | $1.98^{+0.56}_{-0.27}$ |
| $M_V$ (mag) ...... | $3.75 \pm 0.21$ | $3.83 \pm 0.15$ | $3.76 \pm 0.23$ | $3.78 \pm 0.47$ | $3.79 \pm 0.34$ | $3.89 \pm 0.31$ | $4.02 \pm 0.21$ |
| $M_K$ (mag,ESO) . | $2.58 \pm 0.18$ | $2.65 \pm 0.11$ | $2.59 \pm 0.20$ | $2.47 \pm 0.46$ | $2.48 \pm 0.32$ | $2.58 \pm 0.29$ | $2.71 \pm 0.18$ |
| Age (Gyr)....... | $2.0 \pm 0.8$ | $1.7 \pm 0.8$ | $2.0 \pm 0.8$ | $2.4^{+0.7}_{-1.0}$ | $2.5^{+0.7}_{-1.0}$ | $2.3 \pm 0.9$ | $1.8 \pm 0.9$ |
| $A_V$ (mag)....... | $1.900 \pm 0.169$ | $1.895 \pm 0.167$ | $1.897 \pm 0.169$ | $0.831 \pm 0.145$ | $0.832 \pm 0.145$ | $0.829 \pm 0.145$ | $0.822 \pm 0.143$ |
| Distance (pc) ... | $305^{+35}_{-17}$ | $296 \pm 14$ | $303^{+38}_{-18}$ | $296^{+98}_{-33}$ | $296^{+61}_{-29}$ | $281^{+58}_{-20}$ | $266^{+30}_{-15}$ |

[a] Correlation coefficient between the planetary mass $M_p$ and radius $R_p$.
[b] The Safronov number is given by $\Theta = \frac{1}{2}(V_{\rm esc}/V_{\rm orb})^2 = (a/R_p)(M_p/M_\star)$ (see Hansen & Barman 2007).
[c] Incoming flux per unit surface area, averaged over the orbit, measured in units of $10^8\,\mathrm{erg\,s^{-1}\,cm^{-2}}$.

**Table 14**
Model-dependent parameters for outer planets in HAT-P-45 and HAT-P-46

| Parameter | HAT-P-45 **Adopted Model 1 Value** | HAT-P-45 Model 2 Value | HAT-P-45 Model 3 Value | HAT-P-46 Model 1 Value | HAT-P-46 **Adopted Model 2 Value** | HAT-P-46 Model 3 Value | HAT-P-46 Model 4 Value |
|---|---|---|---|---|---|---|---|
| RV and derived parameters for candidate planets HAT-P-45c, HAT-P-46c | | | | | | | |
| $P_c$ (days) | ··· | $15.3 \pm 0.1$ | ··· | ··· | $77.7 \pm 0.6$ | $77.0 \pm 0.4$ | $78.2 \pm 0.3$ |
| $T_{c,c}$ (BJD) [a] | ··· | $2455700.0 \pm 0.3$ | ··· | ··· | $2455695.6 \pm 2.0$ | $2455693.3 \pm 1.7$ | $2455693.1 \pm 0.6$ |
| $T_{14,c}$ (days) [a] | ··· | $0.242 \pm 0.009$ | ··· | ··· | $0.446 \pm 0.058$ | $0.454 \pm 0.088$ | $0.410 \pm 0.030$ |
| $K_c$ ($\mathrm{m\,s^{-1}}$) | ··· | $36 \pm 5$ | ··· | ··· | $81 \pm 9$ | $108 \pm 4$ | $87 \pm 7$ |
| $\sqrt{e}\cos\omega_c$ | ··· | 0 | ··· | ··· | 0 | $0.329^{+0.159}_{-0.243}$ | 0 |
| $\sqrt{e}\sin\omega_c$ | ··· | 0 | ··· | ··· | 0 | $-0.183 \pm 0.193$ | 0 |
| $e\cos\omega_c$ | ··· | 0 | ··· | ··· | 0 | $0.131 \pm 0.116$ | 0 |
| $e\sin\omega_c$ | ··· | 0 | ··· | ··· | 0 | $-0.068 \pm 0.091$ | 0 |
| $e_c$ | ··· | 0 | ··· | ··· | 0 | $0.189 \pm 0.087$ | 0 |
| $\omega_c$ (deg) | ··· | 0 | ··· | ··· | 0 | $302.0 \pm 138.2$ | 0 |
| $M_p \sin i_c$ ($M_J$) | ··· | $0.5 \pm 0.1$ | ··· | ··· | $2.0 \pm 0.3$ | $2.6 \pm 0.1$ | $2.1 \pm 0.2$ |
| $a_c$ (AU) | ··· | $0.130 \pm 0.002$ | ··· | ··· | $0.387^{+0.010}_{-0.007}$ | $0.383^{+0.009}_{-0.006}$ | $0.385 \pm 0.005$ |
| RV and derived parameters for candidate planets HAT-P-45d, HAT-P-46d | | | | | | | |
| $P_d$ (days) | ··· | ··· | ··· | ··· | ··· | ··· | $8.1 \pm 0.1$ |
| $T_{c,d}$ (BJD) [a] | ··· | ··· | ··· | ··· | ··· | ··· | $2455699.0 \pm 0.6$ |
| $T_{14,d}$ (days) [a] | ··· | ··· | ··· | ··· | ··· | ··· | $0.192 \pm 0.014$ |
| $K_d$ ($\mathrm{m\,s^{-1}}$) | ··· | ··· | ··· | ··· | ··· | ··· | $12 \pm 4$ |
| $\sqrt{e}\cos\omega_d$ | ··· | ··· | ··· | ··· | ··· | ··· | 0 |
| $\sqrt{e}\sin\omega_d$ | ··· | ··· | ··· | ··· | ··· | ··· | 0 |
| $e\cos\omega_d$ | ··· | ··· | ··· | ··· | ··· | ··· | 0 |
| $e\sin\omega_d$ | ··· | ··· | ··· | ··· | ··· | ··· | 0 |
| $e_d$ | ··· | ··· | ··· | ··· | ··· | ··· | 0 |
| $\omega_d$ (deg) | ··· | ··· | ··· | ··· | ··· | ··· | 0 |
| $M_p \sin i_d$ ($M_J$) | ··· | ··· | ··· | ··· | ··· | ··· | $0.1 \pm 0.0$ |
| $a_d$ (AU) | ··· | ··· | ··· | ··· | ··· | ··· | $0.085 \pm 0.001$ |

[a] $T_c$: Reference epoch of mid transit that minimizes the correlation with the orbital period. $T_{14}$: total transit duration, time between first to last contact; Barycentric Julian dates (BJD) throughout the paper are calculated from Coordinated Universal Time (UTC).